\documentclass[]{aa} 
\usepackage{psfig,graphics,amssymb} 
\begin{document} 
 
\thesaurus{08.22.3 ; 08.16.4 ; 08.06.3 ; 08.11.1 ; 08.05.3} 
 
\title{Period--Luminosity--Colour distribution and classification  
       of Galactic oxygen--rich LPVs\thanks{Based on data from the  
                                 {\sc Hipparcos} astrometry satellite.}}  
 
\subtitle{I. Luminosity calibrations} 

\author{D. Barth\`es\inst{1}, X. Luri\inst{1}, R. Alvarez\inst{2} and  
M.O. Mennessier\inst{3}}  

 
\institute{Departament d'Astronomia i Meteorologia, Universitat de Barcelona,  
          Avda. Diagonal 647,  E--08028 Barcelona, Spain  
           \and 
           Institut d'Astrophysique, CP 226, Universit\'e Libre de  
           Bruxelles, Bld. du Triomphe, B--1050 Bruxelles, Belgium 
           \and  
           GRAAL (UPRES-A 5024 CNRS), Universit\'e Montpellier II,  
           F--34095 Montpellier Cedex 05, France } 
 
\date { Received 24/02/1999 ; Accepted 08/09/1999 } 
 
\maketitle 

\markboth{D. Barth\`es et al.: PLC distribution and classification of  
Galactic O--rich LPVs --- I.}{} 
 
\begin{abstract} 
 
The absolute $K$ magnitudes and kinematic parameters of about 350  
oxygen--rich Long--Period Variable stars are calibrated, by means of an  
up--to--date maxi\-mum--likelihood method, using {\sc Hipparcos} parallaxes 
and proper motions together with radial velocities and, as additional 
data, periods and $V-K$ colour indices. Four groups, differing by 
their kinematics and mean magnitudes, are found. For each of them, we also 
obtain the distributions of magnitude, period and de-reddened colour of the 
base population, as well as de-biased period--lumi\-nosity--colour relations 
and their two--dimensional projections. The SRa semiregulars do not seem to  
constitute a separate class of LPVs. 
The SRb appear to belong to two populations of different ages. In a PL 
diagram, they constitute two evolutionary sequences towards the Mira  
stage. The Miras of the disk appear to pulsate on a lower--order mode. The 
slopes of their de-biased PL and PC relations are found to be very different 
from the ones of the Oxygen Miras of the LMC. This suggests that a significant 
number of so--called Miras of the LMC are misclassified. This also suggests 
that the Miras of the LMC do not constitute a homogeneous group, but include 
a significant proportion of metal--deficient stars, suggesting a relatively 
smooth star formation history. As a consequence, one may not trivially 
transpose the LMC period--luminosity relation from one galaxy to the other. 
\footnote{Appendix B is only available in electronic form at the CDS via 
anonymous ftp to cdsarc.u-strasbg.fr (130.79.128.5) or via 
http://cdsweb.u-strasbg.fr/Abstract.html} 

\end{abstract} 
 

\section{Introduction} 
 
The {\sc Hipparcos} satellite has provided high--precision parallaxes and 
proper motions of a relatively large number of Long--Period Variable stars 
(LPV) in the solar neighbourhood. In this paper and in the next ones 
(Barth\`es \& Luri \cite{bl99}, Barth\`es et al. \cite{bla99}, 
hereafter Papers II and III), 
which only concern those LPVs belonging to the Asymptotic Giant Branch (i.e. 
Mira, SRa and SRb stars), the {\sc Hipparcos} data are exploited, together 
with radial velocities, $K$ magnitudes, periods, as well as $V-K$ and $J-K$ 
colour indices, by using a specifically adapted maximum--likelihood method of  
luminosity calibration.  
We obtain model distributions of absolute magnitude, dereddened colours and  
period of several groups of stars. We derive de-biased relations between the  
period, the absolute magnitude and the colour indices (PLC relations).  
 
In this paper, the statistical model makes use of the $V-K$ colour index. 
The $J-K$ index is used only {\it a posteriori} in order to check the  
reliability of the results. In paper II, the results will be confronted  
with theoretical models of LPV pulsation. In paper III, a similar work will 
be performed with $J-K$ included in the statistical model.\\

In the next section, the calibration method is presented. The data are  
detailed in Sect. 3. The results of the luminosity calibration are given in  
Sect. 4. Their consequences in terms of PLC, PL, PC and LC relations are  
given and commented in Sect. 5 (these relations concern the sample when they 
involve the $J-K$ index, and the population in all other cases). 
Then, Sect. 6 summarizes and concludes this paper.

\section{Calibration method} 
 
This work is based on the LM method, which has been designed to fully exploit  
the {\sc Hipparcos} data to obtain luminosity calibrations. The mathematical  
foundation of this method was presented in 
Luri (\cite{luri95}) and Luri, Mennessier et al. (\cite{LM96a}). 
Its main characteristics are: 

\begin{itemize} 
\item It is based on a maximum--likelihood algorithm; 
\item It is able to use all the available information on the stars: apparent  
magnitude, galactic coordinates, trigo\-nometric parallax, proper motions,  
radial velocity and other relevant parameters (photometry, metallicity,  
period, etc.), and takes into account, as an additional constraint, the  
existence of mean relations between, e.g., period, luminosity and colour,  
whose analytical form is given {\it a priori}; 
\item It allows a detailed modelling of the kinematics, the spatial 
distribution, and also the distribution of luminosity, period and colour 
of the sample. 

In the implementation presented in this paper, the stars are assumed to be
exponentially distributed about the galactic plane and their velocities to 
follow a Schwa\-rzschild ellipsoid. The period and colour are assumed to 
follow a bivariate normal distribution, including a correlation between the 
two variables. This generates elliptic iso--probability contours in the 
period--colour plane. For each given combination of period and colour, the 
individual absolute magnitudes of the stars are assumed to be 
normal--distributed about the mean value given by a 
period--luminosity--colour relation, e.g. \break 
$M_K = A \, \log P \, + \, B \, (V-K)_0 \, + \, C$. 
The resulting 3D distribution looks like a flattened ellipsoid whose main 
symmetry plane is the PLC relation. All parameters of the model are 
determined by maximum--likelihood estimation; 
\item The method takes into account the observational selection criteria that 
were used when making the sample --- this is very important for obtaining 
unbiased results (Brown et al. \cite{brown97}); 
\item It takes into account the effects of the observational errors; the  
results are not biased by them and even low--accuracy data (which would  
otherwise be useless) can be included; 
\item The galactic rotation is taken into account by introducing in the model 
an Oort--Lindblad first--order differential rotation with $A_0=14.4$, 
$B_0=-12.8\,\, {\rm km}.{\rm s}^{-1}.{\rm kpc}^{-1}$ and $R_{\rm sun} = 
8.5$ kpc; 
\item The interstellar absorption is taken into account, using the 3D model 
of Arenou et al. (\cite{arenou92}). 
\end{itemize} 
\medskip
A further important feature of the LM method is its {\it capability to  
separate and characterize}, in the sample, {\it groups of stars} with 
different properties (e.g. luminosity, kinematics, spatial distribution...). 
The number of groups has to be fixed beforehand (see Sect. 4 for criteria). 
Then, separate results are obtained for each group, and this provides a much 
more meaningful information than a global result for the mixture of all of 
them would.\\ 
 
For the population corresponding to each identified group, the LM method 
provides unbiased estimates of the model parameters, i.e. for the version 
used in this study: 
\begin{itemize} 
\item The parameters of the absolute magnitude distribution, i.e. the  
coefficients of the mean period--luminosity--colour relation, and the 
dispersion around it ($\sigma_{M}$); 
\item The velocity distribution: mean velocities \( (U_{0},V_{0},W_{0}) $  
and dispersions $ (\sigma_{U},\sigma_{V},\sigma_{W}) $; 
\item The spatial distribution: the scale heigth $ Z_{0} $; 
\item The period--colour index distribution: mean of the logarithm of the  
period $ \overline{\log P} $, mean de-reddened colour index, e.g. 
$\overline{(V-K)_{0}}$, the associated dispersions $\sigma_{\log P}$ and, 
e.g., $\sigma_{(V-K)_{0}}$, and the correlation between log period and 
colour; 
\item The percentage of the sample in each group: \%. 
\end{itemize} 
In addition, the parameters of the selection function generating the sample 
are obtained for each group. 

The LM method also yields improved individual distance estimates (and thus 
\emph{improved absolute magnitude estimates}) which take into account all the 
available information on each star: the trigonometric parallax $\pi _{t}$ 
and other measurements (magnitude, $\alpha$, $\delta$, $\mu _{\alpha }$,  
$\mu _{\delta }$, $v_{r}$, $P$, colour). This estimation is free of  
any bias due to observational selection or observational errors,  
because both are taken into account by the method.

\section{Data and other {\it a priori} information} 

\subsection{Sampling}
 
Our sample is made of the 154 Miras and 203 Semiregulars (34 SRa and 169 SRb) 
belonging to the {\sc Hipparcos} Catalogue and for which mean values of both 
$V$ and $K$ magnitudes could be estimated. Their list is given in the 
Appendix B. For 257 stars, $J$ was available too.\\ 

The selection of the LPVs to be included in the {\sc Hipparcos} Input 
Catalogue (Mennessier \& Baglin \cite{mom88}), and thus to be observed by 
the satellite, was based on the General Catalogue of Variable Stars [GCVS] 
(Kholopov et al. \cite{gcvs85}, 1987) and on a criterium of visibility: only 
those stars that were visible (i.e. with an apparent magnitude below the 
{\sc Hipparcos} magnitude limit, $m<m_{lim}$) more than 80\,\% of the time 
were included in the observation programme. This condition can be written as: 
$$ \frac{m_{lim}-m_{max}}{m_{min}-m_{max}}>0.8\,, $$
\noindent translating into a linear relationship $m_{min}<a+b\; m_{max}$. 
On the other hand, the amplitudes of the LPV stars lie within a certain range 
$A_{min}\leq A\leq A_{max}$. One can easily see in Fig.~1 that, with these 
criteria, all LPVs up to a certain magnitude $m_{c}$ are selected and then, 
from $m_{c}$ up to a limiting magnitude $m_{lim}'$, the probability of 
selecting a star decreases linearly. 

As said above, within the frame of the {\sc Hipparcos} Catalogue, our sample 
only includes stars for which mean values of both $V$ and $K$ could be 
obtained. Thus, in any case, the only relevant selection effects (within the 
general frame of the GCVS) are related to the apparent magnitudes of the 
stars. In order to account for these combined effects, a selection function 
$S(m)$ was introduced into the statistical model. Consistently with Fig.~1, 
it was defined so that all stars are selected up to a magnitude 
${\frak m}_{\frak c}$ and then, up to a limiting magnitude 
${\frak m}_{\frak{lim}}'$, the number of selected stars linearly 
decreases. The value of ${\frak m}_{\frak{lim}}'$ was taken equal to the 
apparent magnitude of the faintest star of the sample, and 
${\frak m}_{\frak c}$ is determined (together with all other free parameters) 
by the LM method. In this way, the selection function adapts itself to the 
sample (and to each group that it contains, if several populations are 
assumed).

One must however remember that, despite the relatively large magnitude 
limit of the GCVS ($V \simeq 15$, to be compared to the {\sc Hipparcos} 
limit $\simeq 13$), the 
sample of Mira, SRa and SRb stars found therein is not necessarily complete 
at much lower magnitudes. Indeed, in case of poor data (a frequent problem 
with Semiregulars, according to Lebzelter et al. [\cite{lebz95}]), it is 
difficult to 
detect the variability and to evaluate the amplitude and irregularity of the 
lightcurve. Then, stars may be missing in the GCVS, or SRa and SRb stars 
may be mistaken for each other or for Miras. There is also a significant 
probability to classify an SRa/b star as SR (no identified sub-type) or Lb 
(irregular variable), which two types were excluded from our study before 
applying the magnitude--based selection. On the other hand, due to their 
large amplitude and regularity, Miras are better identified ; in the worst 
case, a Mira is simply mistaken for an SRb, but does not disappear from the 
sample. Summarizing, the boundaries of the three variability types considered 
in this study are more or less blurred, and the used GCVS sample is expected 
to be incomplete, especially concerning Semiregulars. In the previous 
edition of the catalogue, this had spectacular effects: the number of SRb 
stars dropped at $V\simeq 11$, instead of 15 for most other (sub-)types, 
including Mira, SRa and SR (Howell \cite{howell82}). Since then, however, the 
classification has sometimes been revised and many stars have been added. 
As far as we know, the actual incompleteness of the last edition of the GCVS 
has not yet been assessed. Nevertheless, one guesses that the probability of a 
star to have been insufficiently observed mainly depends on the apparent 
magnitudes at max and min and on the period (thus on the mean absolute 
magnitude). As a consequence, the magnitude--based, automatically adjusting 
selection function used in our statistical modelling should account for at 
least a significant part of the sampling bias introduced by the GCVS.

\begin{figure} 
\vspace{3in} 
\includegraphics{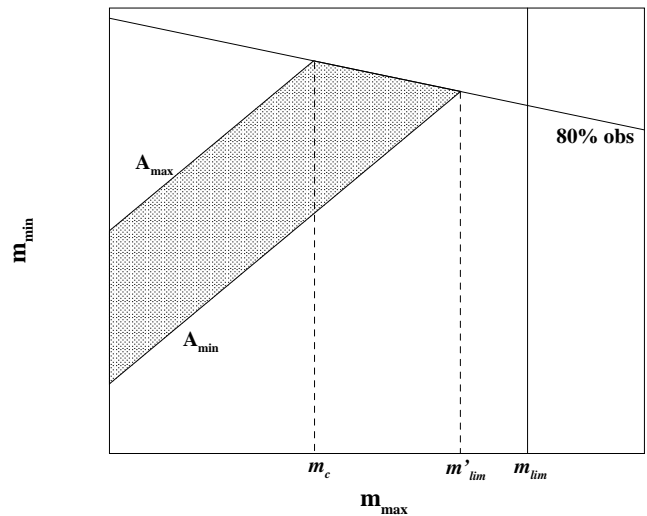} 
\caption{Principle of the selection function of the {\sc Hipparcos} Input 
Catalogue (see text). The selected stars are located within the grey area.} 
\label{fig1} 
\end{figure}

\subsection{Astrometric data} 

For every star of the sample, the coordinates, the parallax and the proper 
motion were found in the {\sc Hipparcos} Catalogue (ESA \cite{esa97}). The 
parallax is negative for 48 Miras, 6 SRa and 8 SRb, but the LM algorithm is, 
by design, able to handle and exploit it. 

For 309 stars, radial velocities were found in the {\sc Hipparcos} Input 
Catalogue [HIC] (Turon et al. \cite{turon92}). Only 23 Miras, 3 SRa and 22 
SRb have no RV data.

\subsection{Magnitudes}

The photometric data that we have chosen are $V$ (represented by visual  
measurements in this study), $J$ and $K$ magnitudes. $K$ was chosen because,  
for LPVs, its behaviour mimics relatively well the one of the bolometric  
magnitude. The $V-K$ colour is much more sensitive to the effective  
temperature and metallicity than $J-K$. On the other hand, the latter colour  
index is less affected by the presence of circumstellar dust shells, and  
it has the advantage that PLC relations using it have already been  
determined for the LMC. 
 
Simple simulations have shown that, for LPV light\-curves with realistic 
amplitudes, periods and asymmetries, the mean magnitude differs from the 
mid--point value (average of the magnitudes at maximum and minimum brightness) 
by at most a few $10^{-1}$ in $V$ and a few $10^{-2}$ in $K$. We will thus 
use indifferently any of these two definitions in this study --- actually the 
mid--point value for $V$ and the mean for $K$. Concerning the latter, it 
is worth noting that it also lies within less than 0.1 of the magnitude 
corresponding to the mean $K$ flux.\\ 

For most Miras and for 10 Semiregulars, the adopted visual magnitudes at  
maximum and minimum light are mean values calculated by Boughaleb 
(\cite{boughaleb95}) from AAVSO data covering 75 years (see Mennessier et 
al. \cite{mom97}). For 5 Miras, mean values of the max and min were deduced 
from AAVSO observations made during the whole {\sc Hipparcos} mission. 
For 5 SR's, we used means at max and min derived from the last 3 decades of 
AAVSO data. 

For 26 Semiregulars, we adopted the mean $V$ magnitudes computed over decades 
by Kiss et al. (\cite{kiss99}), using the Fourier transform. 

For the remaining 28 Miras and for most of the Semiregulars, the visual 
magnitudes at max and min are the ones given by the HIC. 

We remind that the magnitudes at maximum and minimum brightness given by 
the {\sc Hipparcos} Input Catalogue are either averages over decades, found 
in Campbell (\cite{campbell55}), or else estimated means derived from the 
GCVS (Kholopov et al. \cite{gcvs85}, 1987). In the latter case, a statistical 
correction was applied to the catalogue values (and 1.5 mag subtracted in 
case of photographic magnitudes), as explained in the introduction of the HIC. 
For 4 Miras and 2 SRa's for which the HIC magnitudes were adopted, we were 
able to check their consistency (within 0.1 mag) with the 25--year means 
published by the AAVSO (\cite{aavso86}). 

The error bars of visual observations range from $\sigma=0.1$ to 0.5 mag 
according to the brightness. After binning and averaging, the precision 
at maxima is thus better than 0.1 mag; at minima, it may be 
worse. The derived mean magnitude is thus precise within about 0.2 mag. 
However, the uncertainty is larger for the mean maxima and minima derived 
from the GCVS extreme values: $\sigma=0.3$--0.4 mag according to our 
checking. Last, the error bars of the mean magnitudes derived by Kiss et 
al. (\cite{kiss99}) are, of course, negligible compared to the former ones. 
We may thus state that the overall precision of the mean visual magnitudes 
used in this study is about 0.2 mag for Miras and 0.2--0.4 mag for 
Semiregulars.\\

$J$ and $K$ magnitudes (with individual error bars of few $10^{-2}$ mag) 
were found in the Catalogue of Infrared Observations (Gezari et al. 
\cite{gezari96}) --- which includes the large set of JHKL measurements of 
LPVs by Catchpole et al. (\cite{catchpo79}) and the measurements by Fouqu\'e 
et al. (\cite{fouque92}) --- and in recent papers: Groenewegen et al. 
(\cite{groene93}), Guglielmo et al. (\cite{gugliel93}), Whitelock et al. 
(\cite{whitelo94}), Kerschbaum \& Hron (\cite{kersch94}) and Kerschbaum 
(\cite{kersch95}). The number of available data points per star ranges 
from 1 to more than 10, with an average of 1.5 for Miras and 2.2 for 
Semiregulars. As a consequence, considering the overall amplitude, which is 
usually $\la 1$ mag but may reach 1.5 mag for Miras, the error bars 
($\sigma$) of the mean magnitude are a few $10^{-1}$ mag.\\

The mean colour indices $V-K$ and $J-K$ used in this study are the  
differences of the above defined mean magnitudes. The error bars are thus 
roughly 0.5 for the former and, since $J$ and $K$ measurements are usually 
made at the same phase, 0.05 mag for the latter. 

\begin{figure} 
\vspace{3in} 
\includegraphics{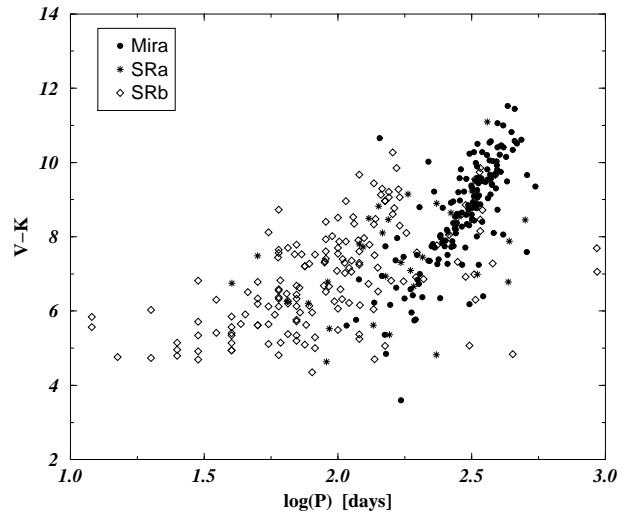} 
\caption{Mean $V-K$ colour versus period of the sample stars (raw data).}  
\label{fig2} 
\end{figure}

\subsection{Periods} 

For 26 Semiregulars, mean periods computed over decades were taken from Kiss 
et al. (\cite{kiss99}). For 21 other SR's, the periods were computed over 
tens of cycles by Bedding \& Zijlstra (\cite{bedding98}), Mattei et al. 
(\cite{mattei97}), Percy et al. (\cite{percy96}) and Cristian et al. 
(\cite{cristian95}). 

For Miras and for the other Semiregulars, the adopted periods are the ones 
given by the HIC. Everytime possible (i.e. for nearly all Miras and for 10 
SR's), we have checked that they are very close to the 75--year means 
calculated by Boughaleb (\cite{boughaleb95}) from AAVSO data covering 75 
years; the differences of a very few \% correspond to the cycle--to--cycle 
fluctuations (see Mennessier et al. \cite{mom97}). For 4 Miras and 2 SRa's, 
we were able to check that the HIC periods lie within 1--2\,\% of the 
25--year means published by the AAVSO (\cite{aavso86}). Concerning the other 
stars, we can only guess the overall quality of the HIC by checking all stars 
used by Kiss et al. (\cite{kiss99}), Bedding \& Zijlstra (\cite{bedding98}), 
Mattei et al. (\cite{mattei97}), Percy et al. (\cite{percy96}) and 
Cristian et al. (\cite{cristian95}): for SRa stars, only 4\,\% are found 
spurious (error $\ga 10\,\%$) and 83\,\% are very good (error $\la 3\,\%$); 
for SRb stars, about 25\,\% of the HIC periods appear spurious and 66\,\% 
very good. 

As a consequence, about 15\,\% of the periods may be spurious in the sample of 
Semiregulars used in this paper.

\subsection{Constraints}

In addition to these individual data, it is known that O--rich LPVs in the  
LMC follow linear mean relations between the absolute magnitude  
($M_K$ or $M_{\rm bol}$) and the logarithm of the period, and also  
near--infrared colour indices such as $(J-K)_0$ (Feast et al. \cite{feast89},  
Hughes \& Wood \cite{hughes90}, Hughes \cite{hughes93}, Wood \& Sebo 
\cite{wood96}, Kanbur et al. \cite{kanbur97}, Bedding \& Zijlstra 
\cite{bedding98}).  
The existence of a linear \{$M_I$, $\log P$\} relation has also been shown  
(Feast et al. \cite{feast89}, Pierce \& Crabtree \cite{pierce93}). Moreover, 
Alvarez et al. (\cite{alvarez97}), applying to {\sc Hipparcos} data an early 
version of the LM method that {\it does not} assume the existence of any PL 
or PLC relation, have shown that Oxygen--rich Miras in the solar 
neighbourhood do follow linear \{$\log P$, $M_K$\} relations. 
On the other hand, consistent with Kerschbaum \& Hron (\cite{kersch92}), a 
simple plot of our raw data (see Fig. 2) strongly suggests that Miras and 
Semiregulars are distributed around at least two linear \{$V-K$, $\log P$\} 
relations, the one of Miras being peculiarly well--defined.

As a consequence, the calibrations presented in this series of papers have  
been performed under the assumption (constraint) that there exist in the  
sample such PLC relations whose de-biased coefficients are to be calculated  
by the algorithm. The validity of this choice is confirmed by the consistency 
of the so--derived luminosities with the ones found without making this 
assumption (Mennessier et al. \cite{mom99}). 

\begin{figure} 
\vspace{3in} 
\includegraphics{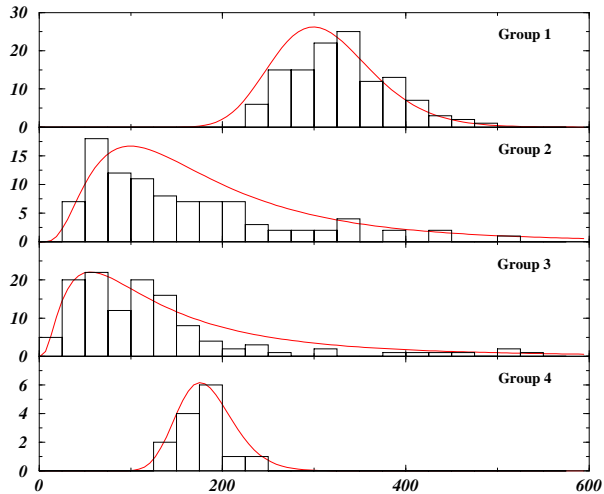} 
\caption{De-biased model period distributions (lines) and histograms  
of the data} 
\label{fig3} 
\end{figure}

\begin{table*} 
\caption[]{Model parameters of the four groups ($\theta$ denotes the fitted  
values and $\sigma$ their uncertainties)} 
\begin{flushleft} 
\begin{center} 
\begin{tabular}{lrrrrrrrr} 
\hline 
  & \multicolumn{2}{r}{Group 1} &\multicolumn{2}{r}{Group 2}& \multicolumn{2}{r}{Group 3}  & 
\multicolumn{2}{r}{Group 4} \\ \hline 
  & $\theta $ & $\sigma$  &  $\theta $ & $\sigma$  & $\theta $ & $\sigma$ & $\theta $ & $\sigma$ \\  \hline 
 $U_0$  [km/s]  & -11.4  & 3.8 & -33.8  & 5.9 &  -3.9  & 5.4 & -33.1 & 66.2\\ 
$\sigma_U$      &  43.3  & 5.2 &  48.1  & 4.9 &  34.6  & 3.5 & 145.3 & 36.3\\  
 $V_0$   [km/s] & -31.9  & 5.0 & -46.4  & 4.4 & -19.1  & 1.8 &-178.6 & 37.2\\ 
$\sigma_V$      &  29.8  & 2.0 &  37.9  & 4.5 &  17.5  & 3.3 & 102.4 & 26.3\\ 
 $W_0$  [km/s]  & -11.5  & 3.6 & -10.8  & 5.1 &  -9.3  & 2.0 &  -4.8 & 30.6\\  
$\sigma_W$      &  27.0  & 2.5 &  38.9  & 4.5 &  14.1  & 1.8 &  70.0 & 15.4\\  
$Z_0$ [pc]      &  368   & 55  & 476    & 54  & 174    & 28  &       &  \\  
\hline 
$\overline{\log P}$ & 2.48 & 0.01 & 2.00 & 0.06 & 1.75 & 0.06 & 2.24 & 0.03\\ 
$\sigma_{\log P}$   & 0.04 & 0.01 & 0.27 & 0.03 & 0.29 & 0.02 & 0.06 & 0.01 \\ 
$\overline{(V-K)_0}$ & 8.47 & 0.11 & 5.75 & 0.13 & 6.19 & 0.28 & 5.56 & 0.46\\ 
$\sigma_{(V-K)_0}$  & 0.52 & 0.04 & 1.04 & 0.09 & 1.14 & 0.08 & 0.86 & 0.17\\ 
Cor.$(\log P,(V-K)_0)$&-0.85 &0.04 &-0.46 & 0.07 &-0.65 & 0.06 &-0.49 & 0.44\\ 
\hline  
${\frak m}_{\frak c}$ & -4.0 & 2.4 & -2.7 & 1.6 & -4.0 & 1.3 & 1.8 & 1.5 \\
\hline
 \%                 & 31.8 & 0.02 & 29.3 & 0.03 & 34.9 & 0.03 &  3.9 & 0.02\\ 
\hline 
\end{tabular} 
\end{center} 
\end{flushleft} 
\label{tab1} 
\end{table*}

\section{Calibration and classification}

The assumed number of model populations ({\it groups}) is constrained by the  
limited number of sample stars (357) and by the number of parameters to be  
fitted (18 per group). Its relevant value may be determined by means of a 
Wilks test (Soubiran et al. \cite{soubiran90}, Wilks \cite{wilks63}). 
This test basically checks the significance of the likelihood increase 
obtained when the number of free parameters (in particular the number of 
groups) is increased. Considering that we are dealing with 2 or 3 types of 
variable stars and 2 or 3 galactic populations (Luri et al. \cite{LM96b}, 
Alvarez et al. \cite{alvarez97}), several computations 
were carried out with 2, 3 and 4 groups. Wilks test indicated that the 
four--groups solution was still significant. Computation with five groups was 
not pursued until convergence because the number of free parameters was 
obviously too high (89 for 357 stars). To this respect, it is worth remarking 
that Group 1 was always clearly separated, while the other groups were mixed 
when less than 4 groups were used. 

The fitted parameters of the model distributions corresponding to the 
4--groups solution are given in Table 1 and in Sect. 5.1, with $1\sigma$ 
error bars derived from Monte Carlo simulations. 
 
The de-biased model distributions of period are shown in Fig. 3.  
 
The tridimensional \{$P$, $M_K$, $(V-K)_0$\} distributions of the  
calibrated data and of the model populations are displayed in Fig. 4.

\begin{figure*}
  {\centering \resizebox*{18.cm}{!}
              {\includegraphics{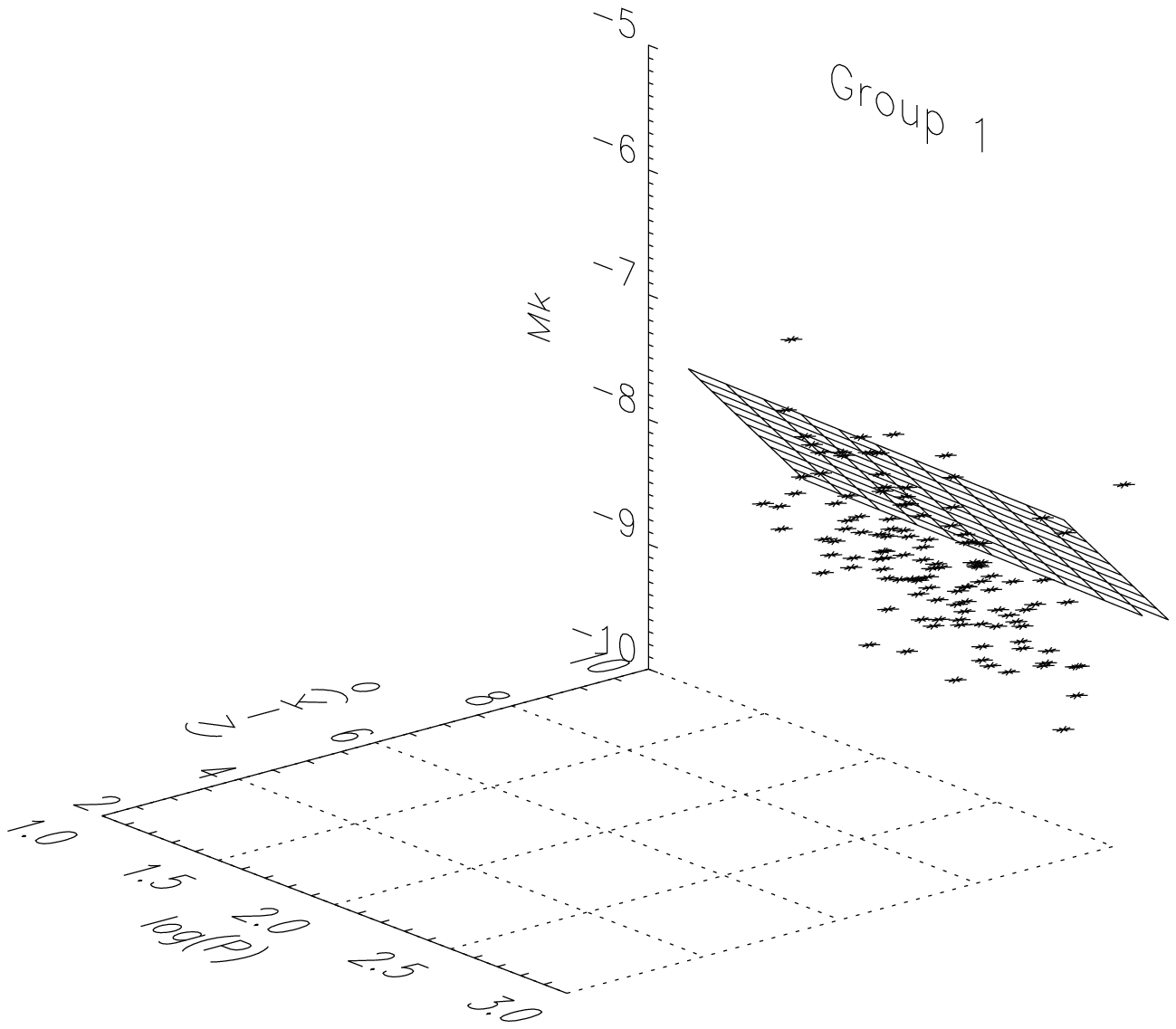}
               \includegraphics{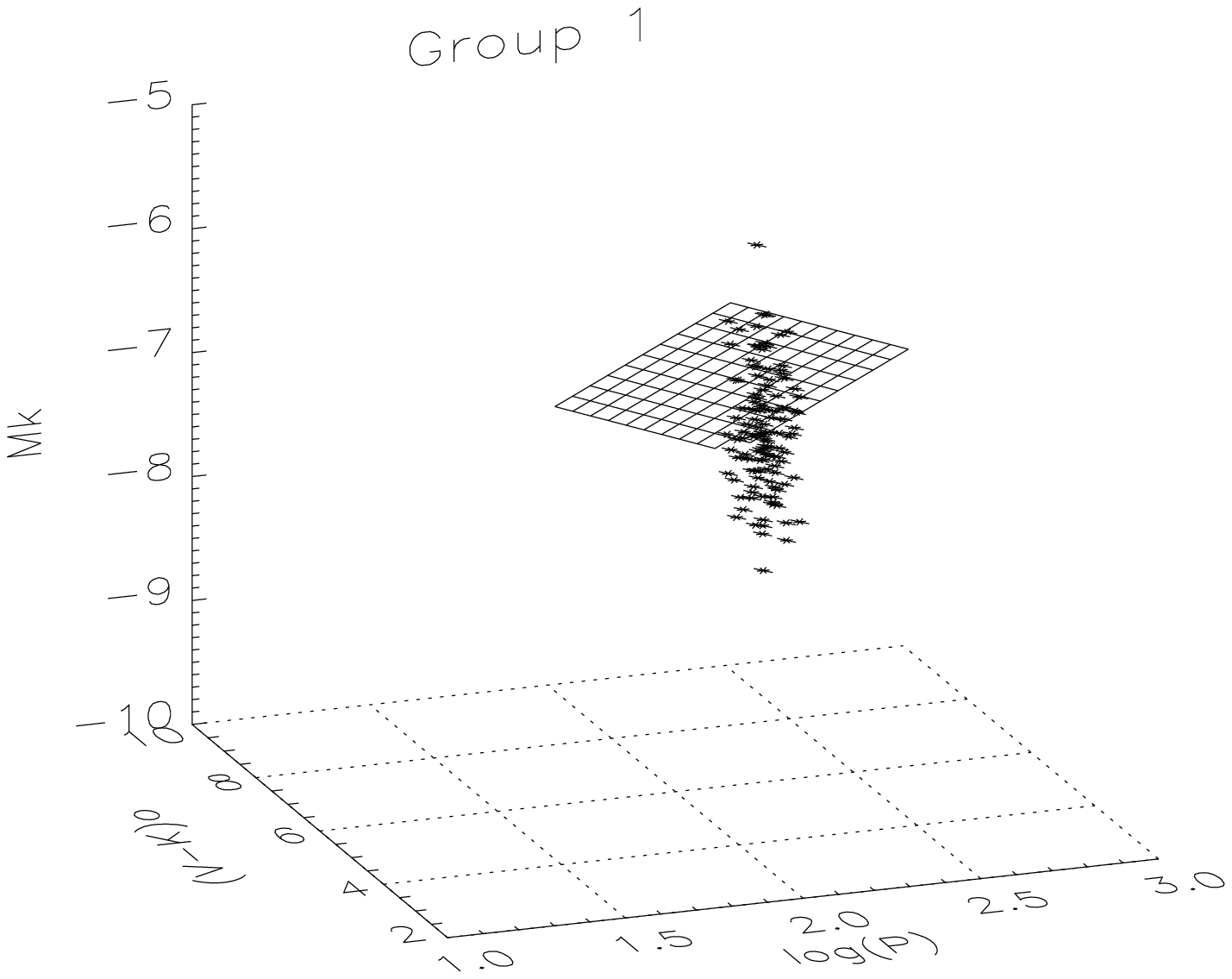}
               \includegraphics{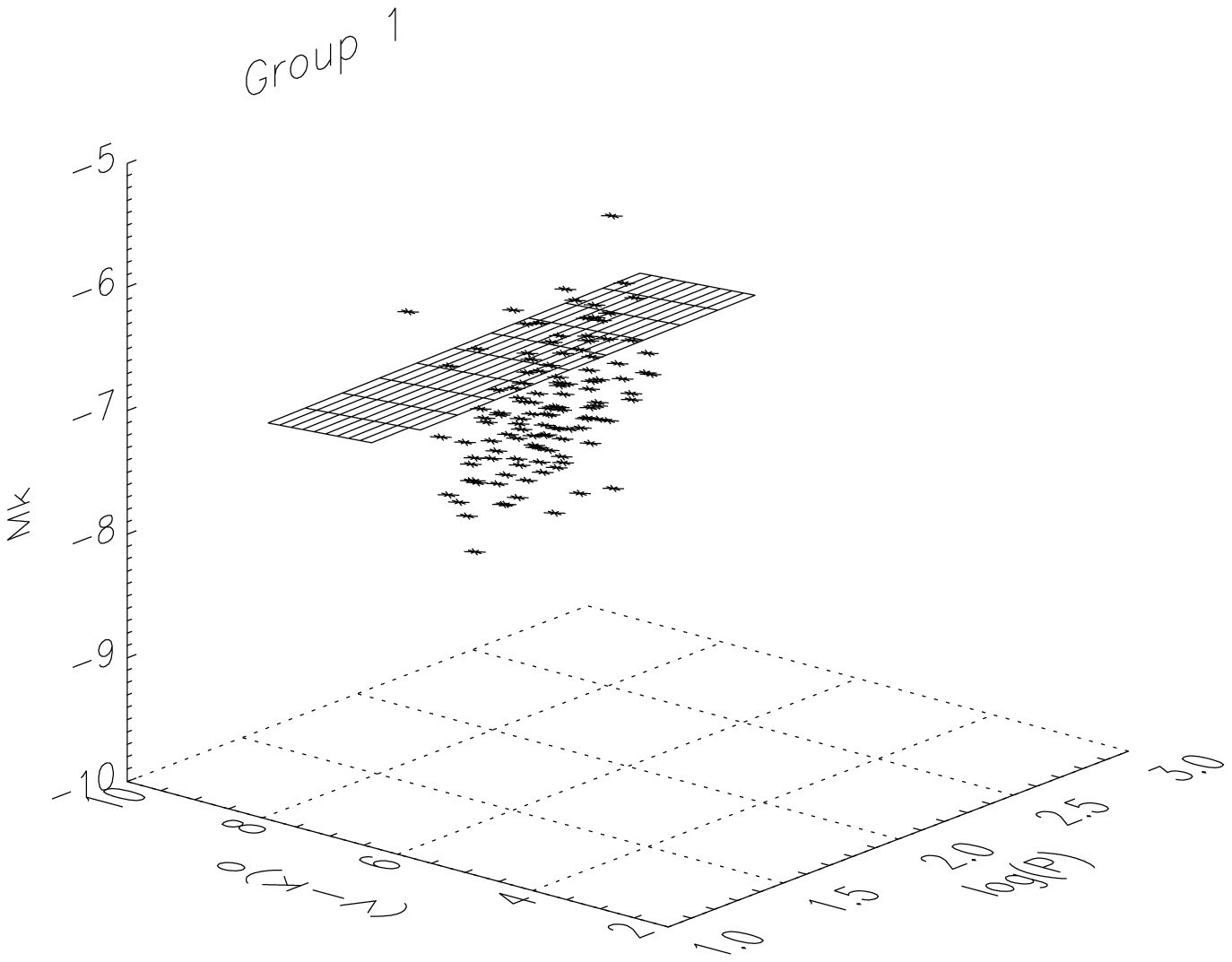}} \par}
  {\centering \resizebox*{18.cm}{!}
              {\includegraphics{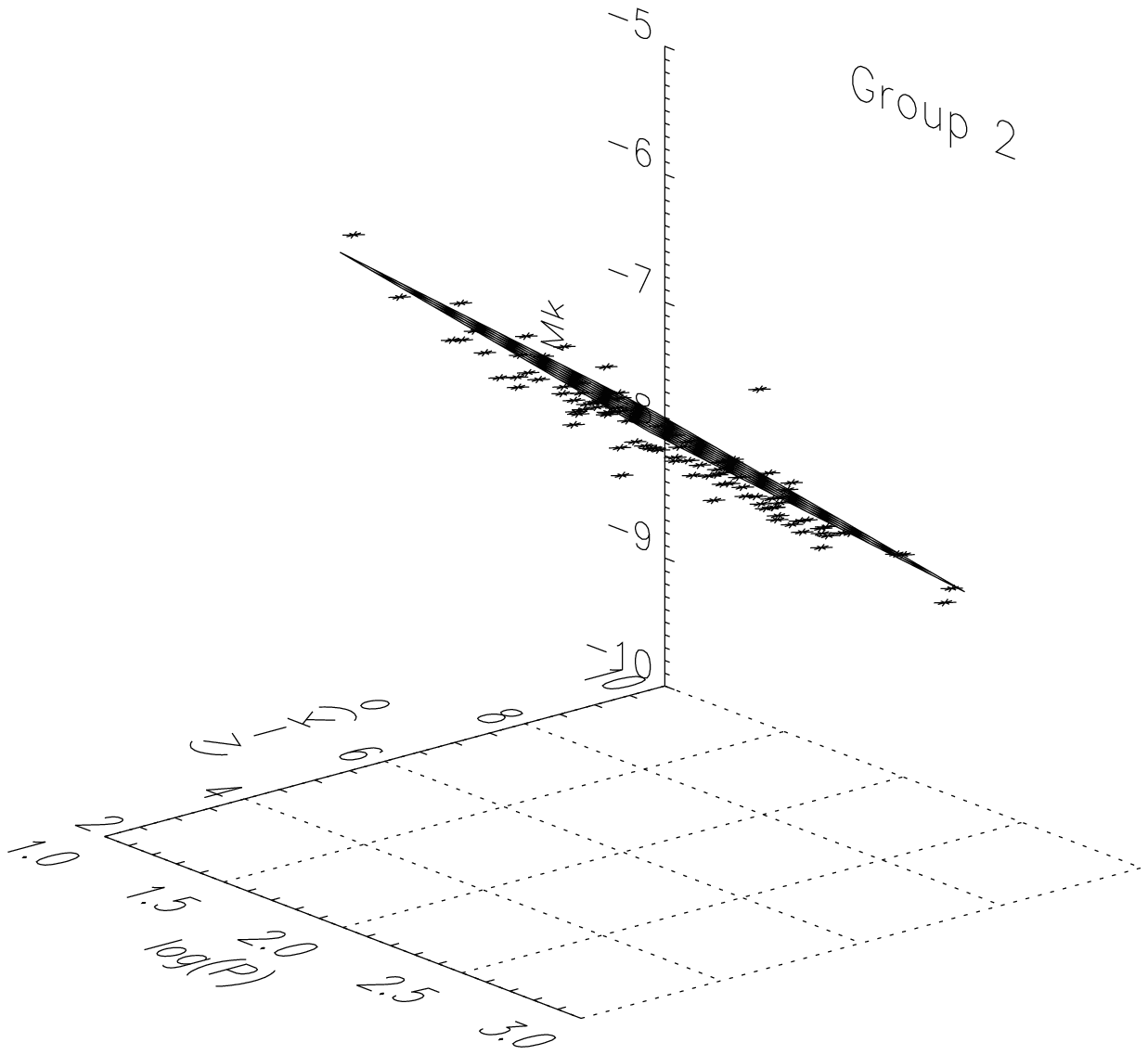}
               \includegraphics{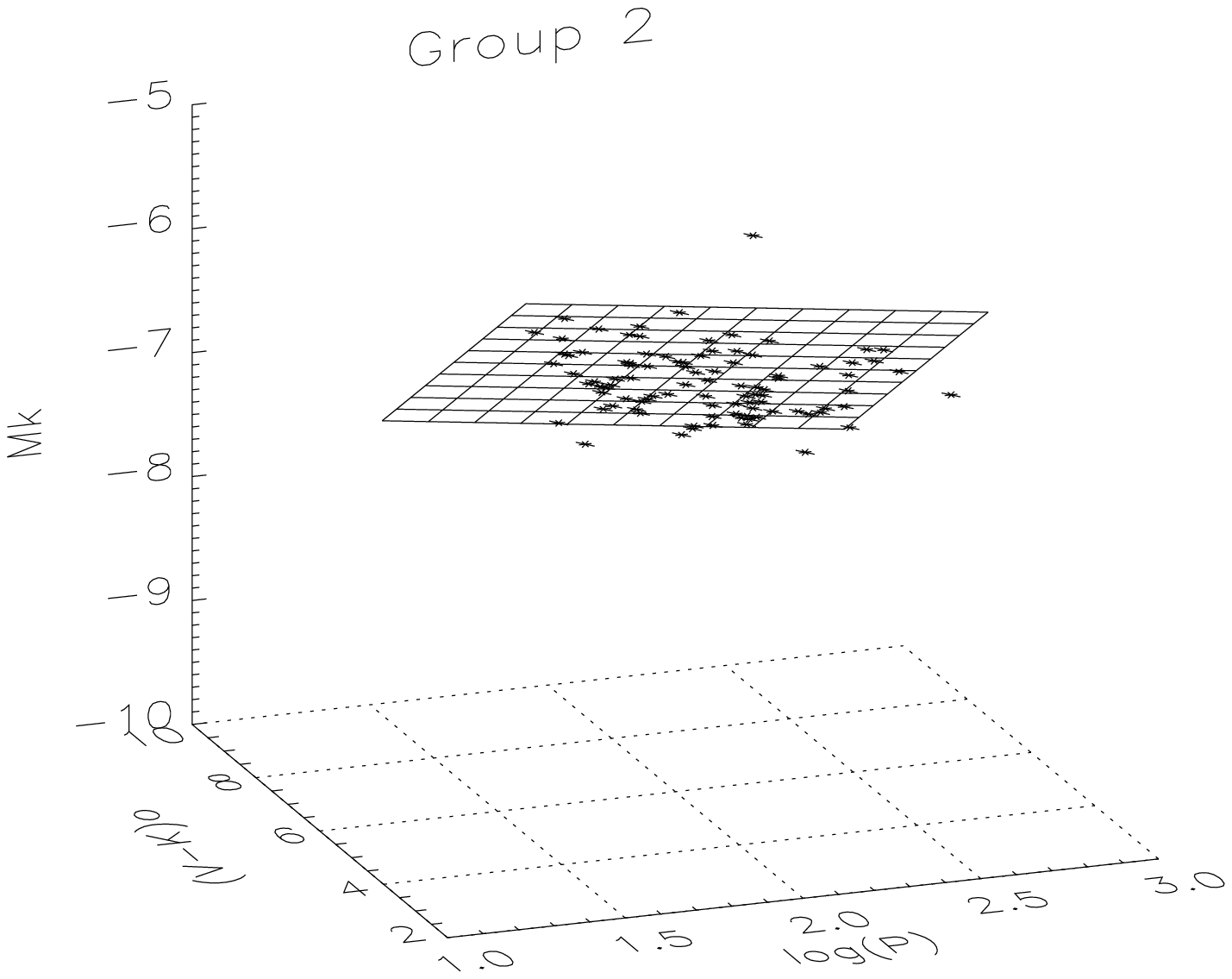}
               \includegraphics{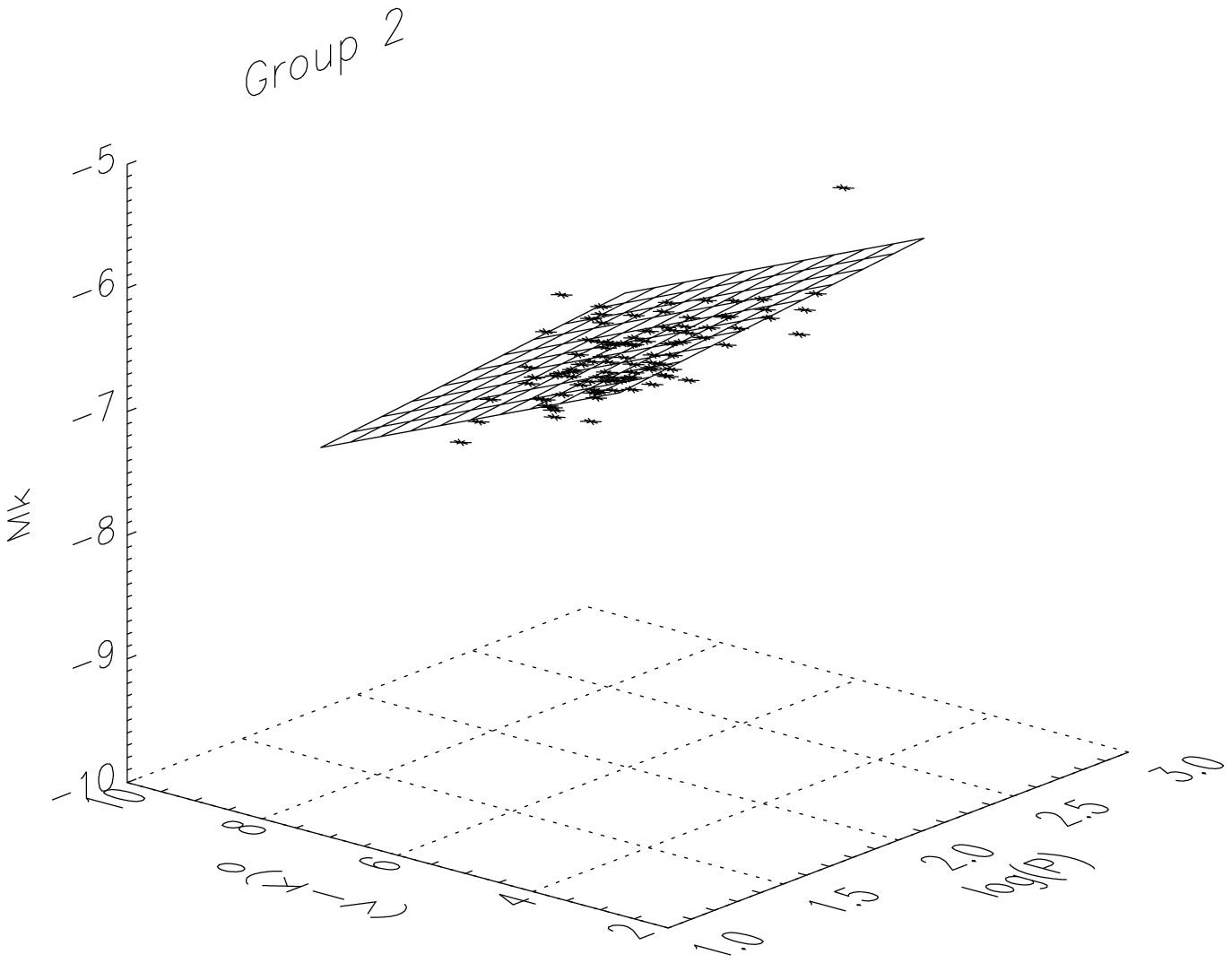}} \par}
  {\centering \resizebox*{18.cm}{!}
              {\includegraphics{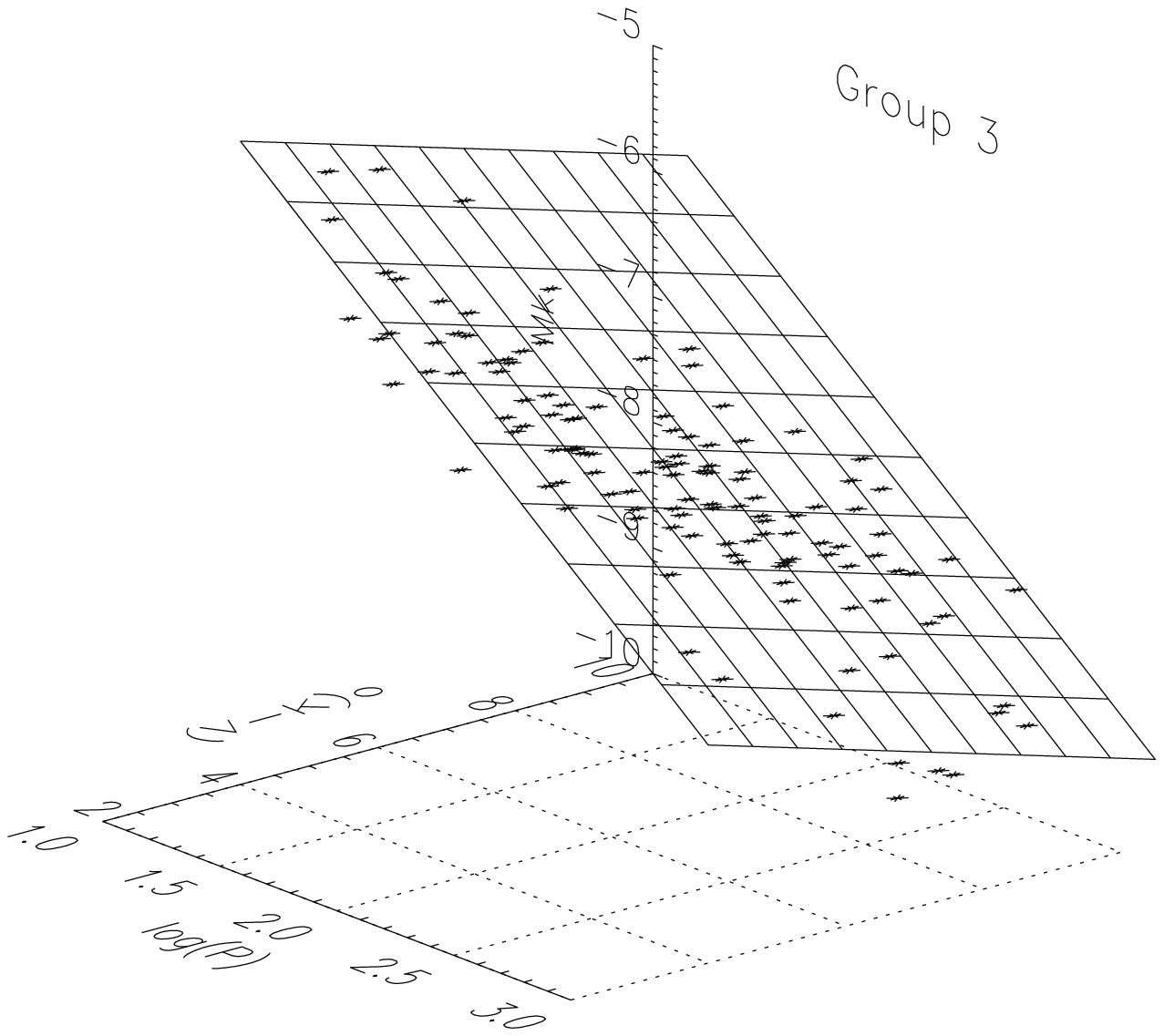}
               \includegraphics{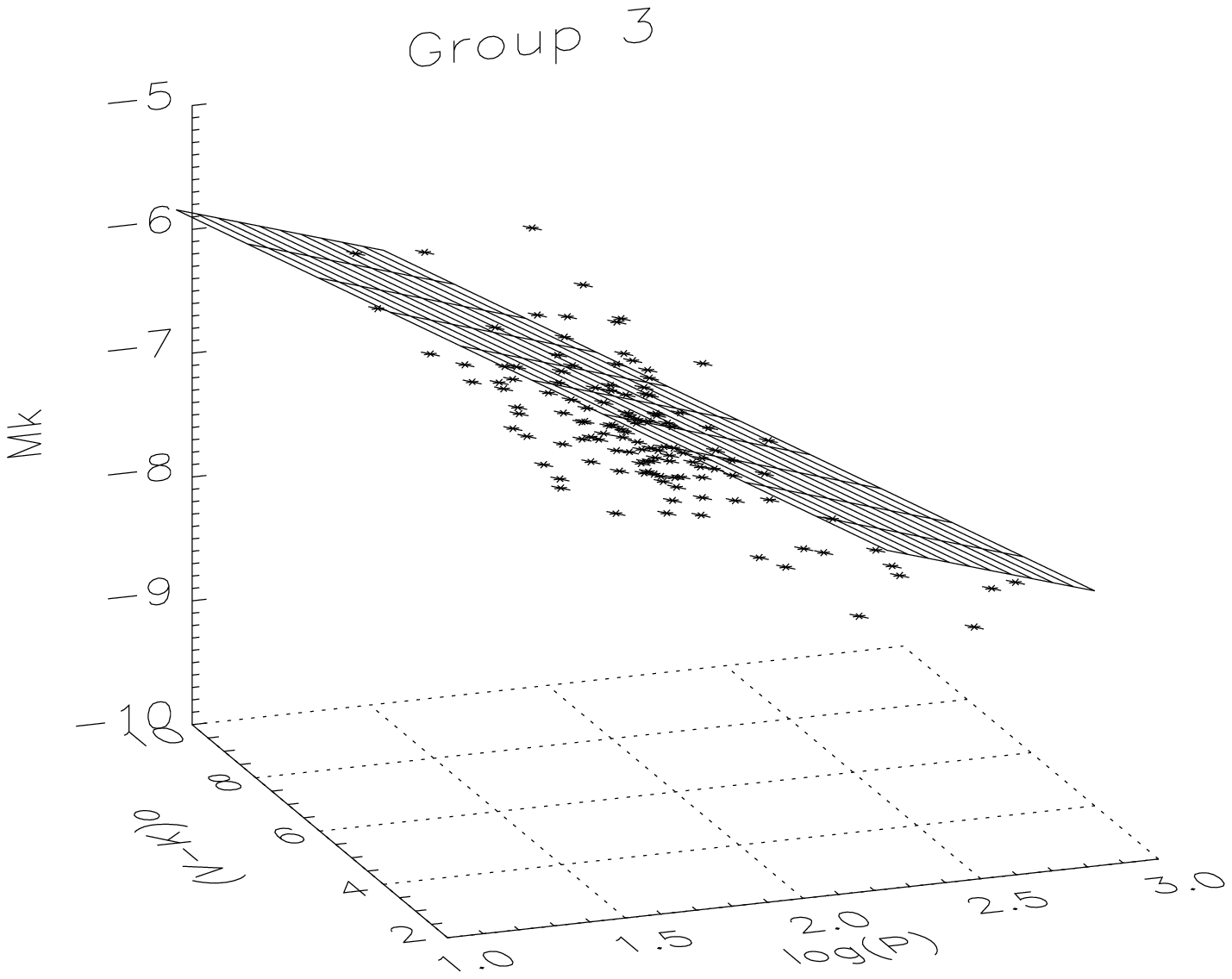}
               \includegraphics{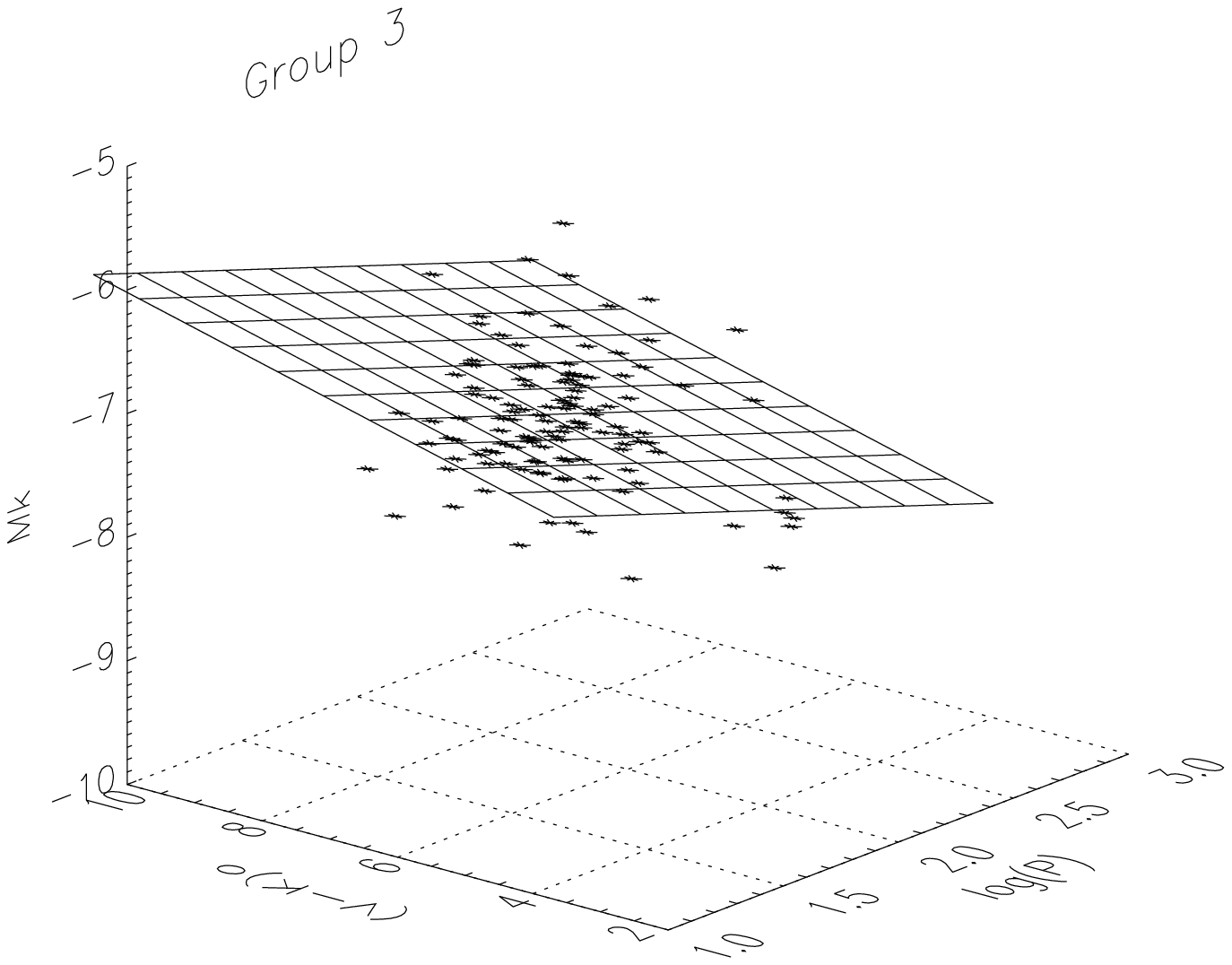}} \par}
  {\centering \resizebox*{18.cm}{!}
              {\includegraphics{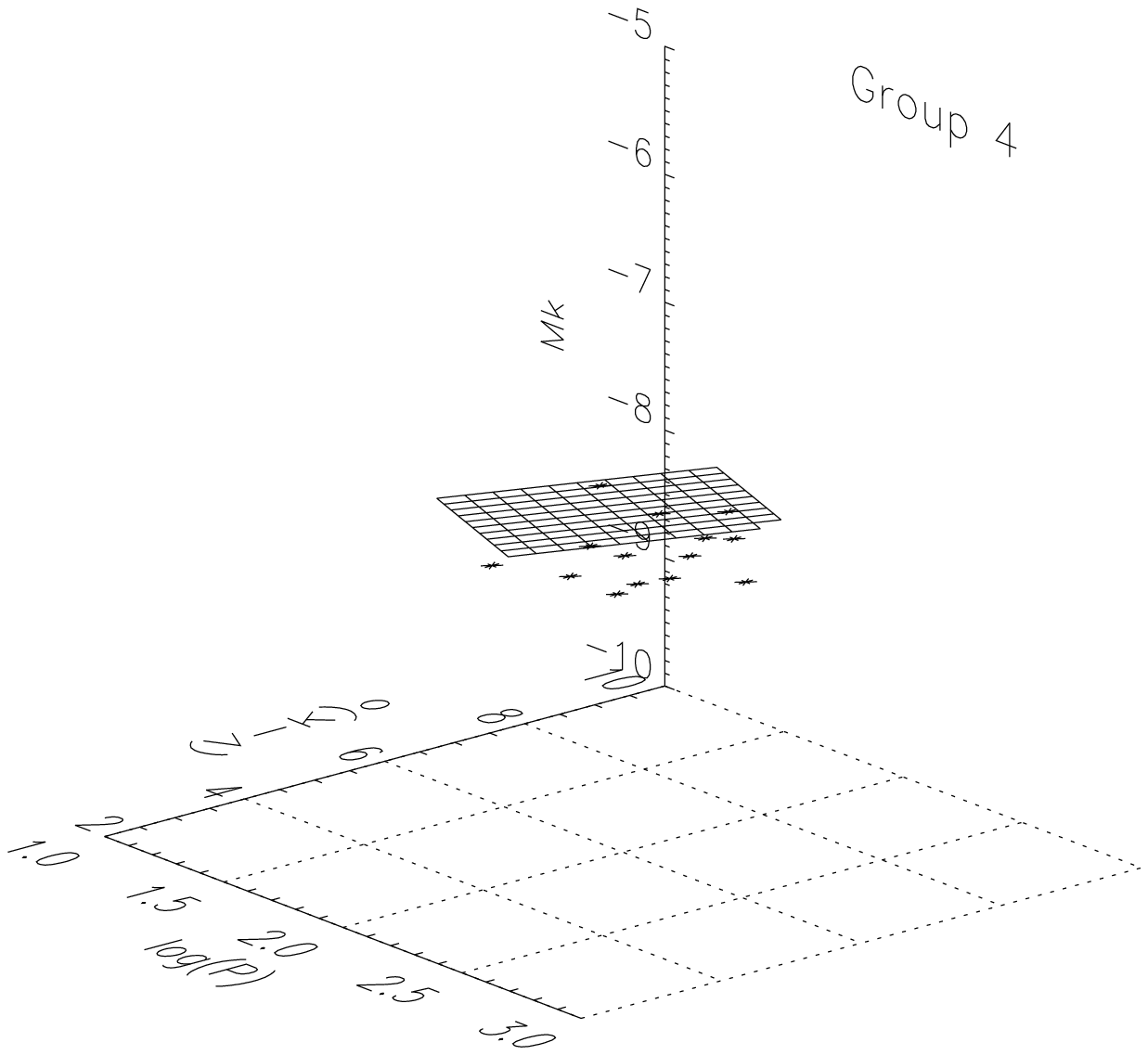}
               \includegraphics{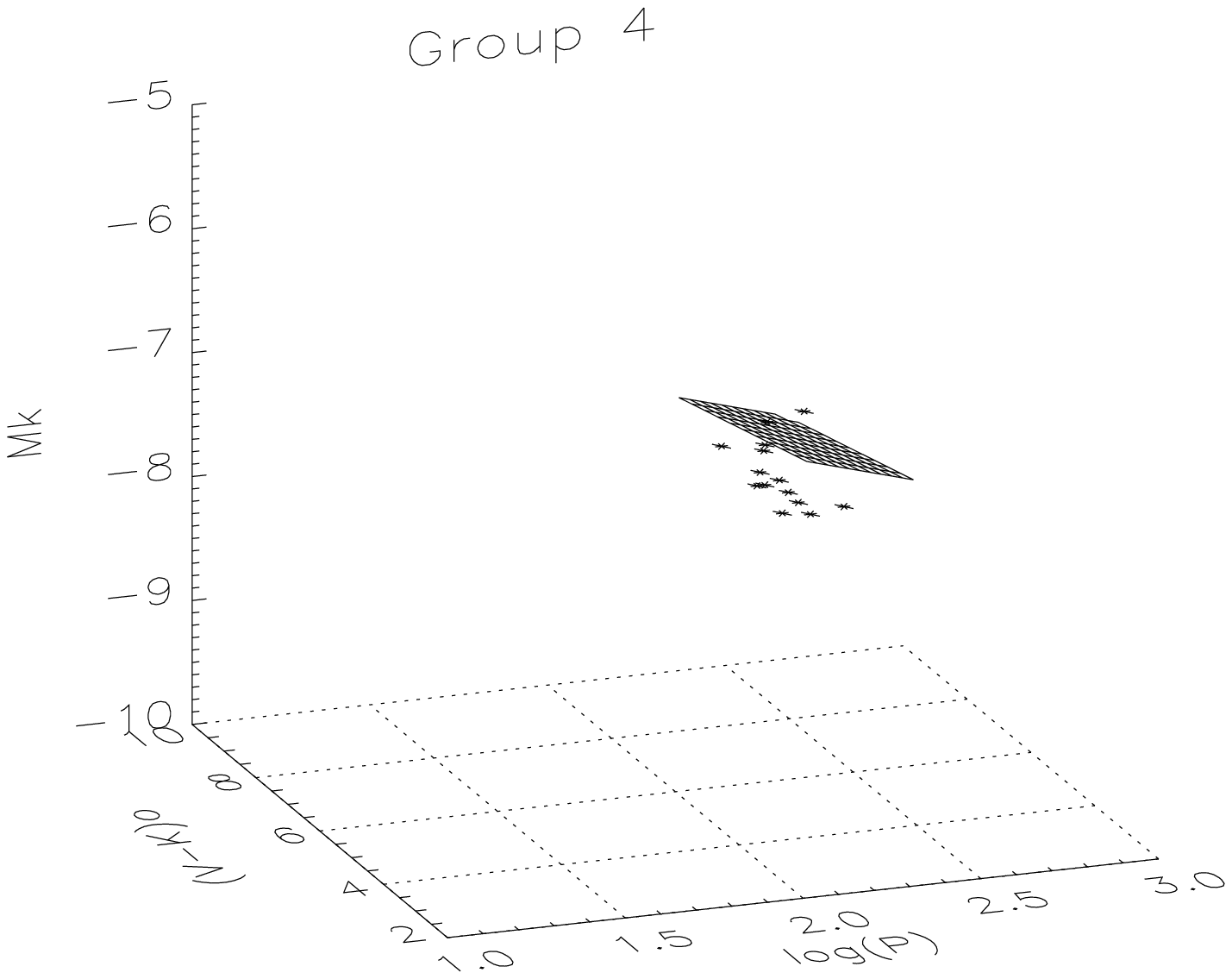}
               \includegraphics{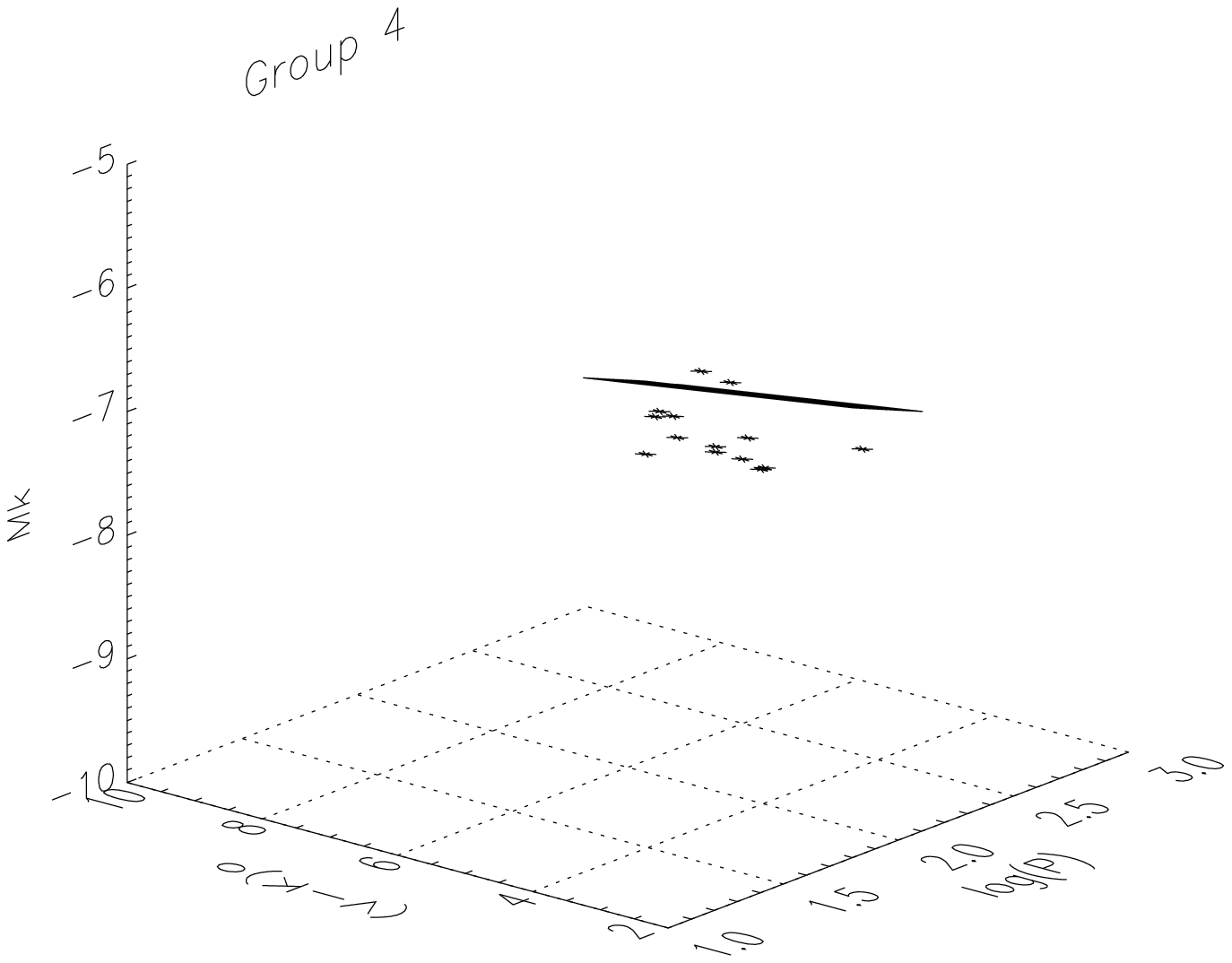}} \par}
\caption{Calibrated period--luminosity--colour distributions: individual 
data and main symmetry plane (i.e. {\it de-biased} PLC relation) of the 
model distribution of each group.} 
\label{fig4} 
\end{figure*}

The groups may get the following interpretation in terms of  
kinematics: 
 
Group 1 (121 stars: 102 Miras, 6 SRa, 13 SRb) and Group 2 (96 stars: 
54 SRb, 26 Miras, 16 SRa) have very similar kinematics, corresponding to 
old disk stars. Group 3 (125 stars: 102 SRb, 12 Miras, 11 SRa) has a  
younger kinematics than the previous ones. Group 4 (14 stars: 13 Miras,  
1 SRa) has the kinematics of extended--disk or halo stars. 
 
One can immediately see that the SRb stars constitute two populations 
of different ages. SRa stars are spread over all groups, with no clear  
``preference''. Most Miras appear in Group 1, with the same kinematics as  
the old SRb population. 
 
\section{Period---Luminosity---Colour relationship} 
 
\subsection{PLC relations} 
 
The fitted distributions of period, magnitude and colour correspond 
to the following PLC relations (where the error bars correspond to 
$\pm 1\sigma$ deviations, as estimated using Monte Carlo simulations):\\ 

\begin{itemize} 
\item{Group 1}:
\medskip
\begin{eqnarray*}
M_K & =  -1.07_{[\pm 0.50]} \log P & - 0.37_{[\pm 0.13]} (V-K)_0 \\
    &                               & - 1.49_{[\pm 0.72]}
\end{eqnarray*}  
$\sigma_M\, =\, 0.63_{[\pm 0.13]}$\\ 
\item{Group 2}: 
\medskip
\begin{eqnarray*}
M_K & =  -0.37_{[\pm 0.94]} \log P & -0.42_{[\pm 0.24]} (V-K)_0\\ 
    &                               & -3.23_{[\pm 1.35]}
\end{eqnarray*}
$\sigma_M\, =\, 0.32_{[\pm 0.17]}$\\ 
\item{Group 3}: 
\medskip
\begin{eqnarray*}
M_K & =  -1.69_{[\pm 0.54]} \log P & -0.16_{[\pm 0.14]} (V-K)_0\\ 
    &                               & -2.53_{[\pm 0.78]} 
\end{eqnarray*}
$\sigma_M\, =\, 0.50_{[\pm 0.08]}$\\ 
\item{Group 4}: 
\medskip
\begin{eqnarray*}
M_K & =  -0.81_{[\pm 1.72]} \log P & -0.09_{[\pm 0.45]} (V-K)_0\\ 
    &                               & -4.76_{[\pm 2.48]}
\end{eqnarray*}
$\sigma_M\, =\, 0.48_{[\pm 0.21]}$\\ 
 \end{itemize} 

The coefficients of the relation found for Group 4 are obviously very  
uncertain. This is not surprising, in view of the small number of stars 
(15), their small dispersion, and the number of parameters to 
estimate. For Groups 1, 2 and 4, the error bars on the zero point are 
relatively large; this is due to the fact that the means of the three 
variables are far from zero, and thus any slope uncertainty rebounds 
magnified on the zero point.

\subsection{Projection onto the \{$P, M_K$\} plane} 
 
The tridimensional model distributions may be projected onto the  
period--luminosity plane. The elliptic--looking lines shown in Fig. 5 
are the projections of the isoprobability contours that, in the mean 
PLC plane, correspond to a $2\sigma$ deviation. The offset between 
the data and the model populations is due to the sampling bias, which 
is suppressed by the LM algorithm (see Appendix A for details). 
Then, Period--luminosity relations, liable to be compared to the ones 
observed in the Magellanic Clouds, are derived by means of a linear 
least--squares fit to the contours. Monte--Carlo simulations, as well as 
analytic computations, have shown that this is equivalent to a fit onto the 
projected population itself. Finally, the error bars of the coefficients 
are estimated, for each group, by applying the standard least--squares 
procedure to a simulated unbiased sample (thus they may be directly compared 
to the error bars usually given for the LMC stars).\\ 

The results are the following: 

\begin{itemize} 
\item {Group 1}:  
$$M_K = -5.04_{[\pm 0.72]} (\log P - 2.48) - 7.26_{[\pm 0.06]}$$  
\item {Group 2}:  
$$M_K = -1.13_{[\pm 0.16]} (\log P - 2.00) - 6.41_{[\pm 0.05]}$$ 
\item {Group 3}:  
$$M_K = -2.11_{[\pm 0.13]} (\log P - 1.75) - 6.51_{[\pm 0.05]}$$ 
\item {Group 4}:  
$$M_K = -1.37_{[\pm 2.32]} (\log P - 2.24) - 7.05_{[\pm 0.14]} $$ 
\end{itemize} 
 
It may be noticed that Groups 2 and 3  have similar mean magnitudes.  
The same holds for Groups 1 and 4.\\ 

\begin{figure} 
\vspace{3in} 
\includegraphics{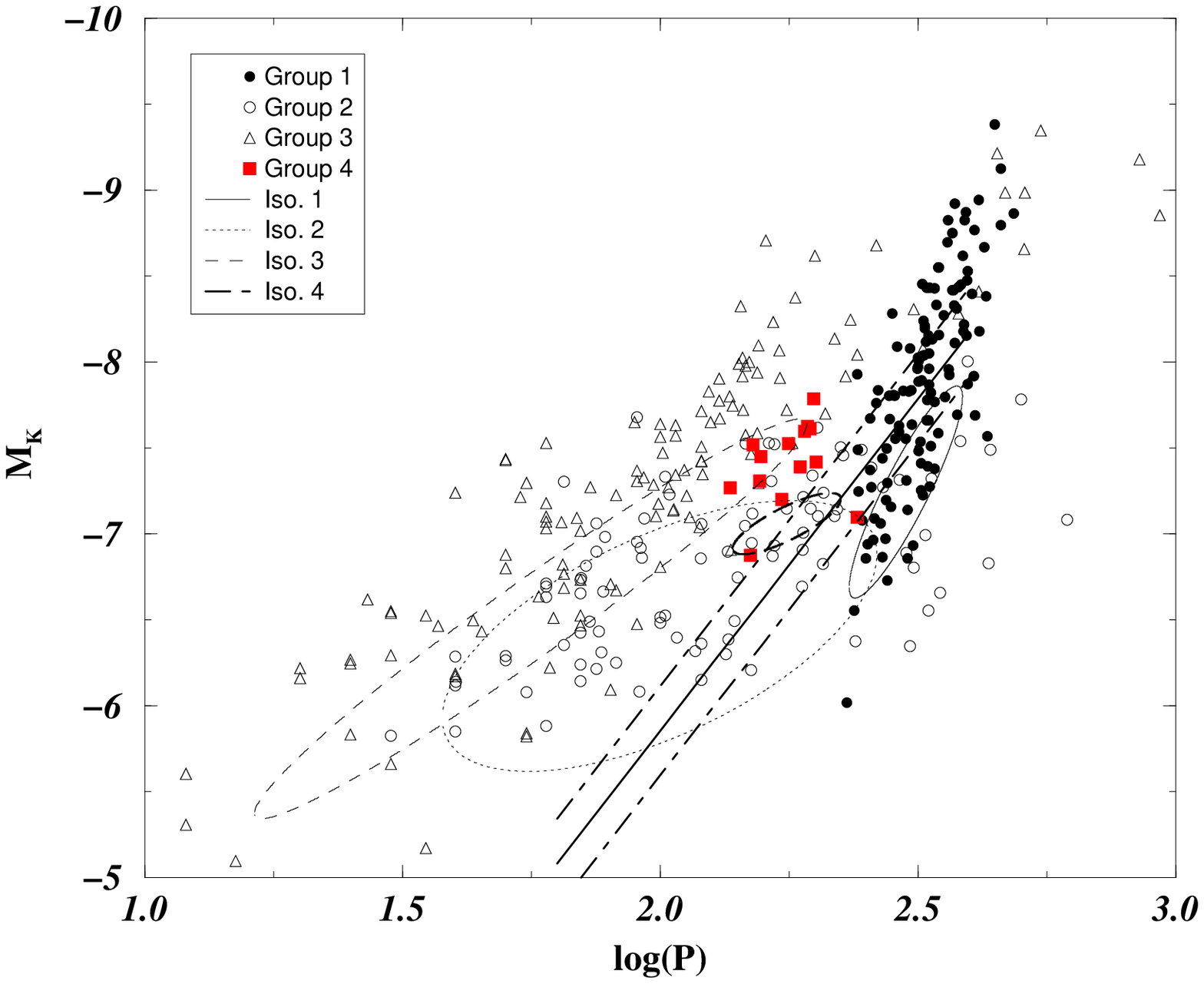} 
\caption[]{ PL calibrated distributions: individual data and projected 
model distributions ($2\sigma$ isoprobability contours in the mean PLC 
plane). The Mira strip of the LMC is also shown (thick lines). } 
\label{fig5} 
\end{figure} 

From a sample of 29 O--rich Miras of the Large Magellanic Cloud with period  
$\le 420$ days, Feast et al. (\cite{feast89}) derived the following relation  
(were the error bars correspond to $1 \sigma$ deviations):  
 
$$M_K = - 3.47_{[\pm 0.19]} \log P + 0.93_{[\pm 0.45]}$$  
$$\sigma_M=0.13$$ 
 
\noindent assuming a distance modulus of 18.55.  
 
Based on a sample of more than a hundred Oxygen--rich Miras of the LMC, the  
solution of Hughes \& Wood (\cite{hughes90}) is, under the same assumptions:  
 
$$M_K = - 3.86_{[\pm 0.18]} (\log P - 2.4) -7.40_{[\pm 0.02]}$$  
$$\sigma_M=0.26$$  
 
From a sample of 79 Miras, unfortunately including a significant number of  
Carbon stars, Hughes (\cite{hughes93}) derived:  
$$M_K = - 3.75_{[\pm 0.14]} (\log P - 2.4) -7.45 _{[\pm 0.02]}$$  
$$\sigma_M=0.13$$ 

Obviously, the slopes are significantly different from that of any Galactic 
PL relation found above. Such a difference was also observed between the LMC 
and Globular Clusters stars by Menzies \& Whitelock (\cite{menzies85}). 
It cannot be due to the well--known steepening of the PL relation at periods 
larger than 420 days, i.e. relatively high mass (Feast et al. 
\cite{feast89}), since only a very few sample stars may be concerned. 
The first possible explanation is that the so--called Miras of the LMC 
include, in fact, a significant number of SRb semiregulars, especially at 
``short'' periods. Indeed, for the outer galaxies, the observers use to 
call ``Miras'' the LPVs with an amplitude larger than a given threshold 
(e.g. 0.9 mag in $I$), corresponding to the maximum amplitude of SRa stars 
according to the GCVS. This criterium is obviously not sufficient and the 
slope of the so--called LMC ``Mira'' strip should then be intermediary 
between our Groups 1 and 2. A second explanation of the slope discrepancy is 
that the shorter--period ``Miras'' in the LMC include a population 
more--or--less equivalent to our Group 4, i.e. metal--deficient with a mean 
mass similar to or lower than the one of the main population. This, too, 
would lead to a shallower global PL relation. The existence of such a 
population had been suggested by Wood et al. (\cite{wood85}) and Hughes et 
al. (\cite{hughes91}). Of course, these two explanations do not exclude 
each other. \\

It is also worth noting that the PL slopes of Groups 2 and 3 are much smaller 
than the one of Group 1 (Miras), and similar to the one of the  
evolutionary tracks ($-1.67$), derived by Bedding \& Zijlstra 
(\cite{bedding98}) from the works of Whitelock (\cite{whitelo86}) and 
Vassiliadis \& Wood (\cite{vassili93}). 
Moreover, in Group 2 as well as in Group 3, the proportion of SRb's (as 
defined by the GCVS) decreases towards longer periods: at $P> 200$ days, 
they represent less than 25\,\% of the stars of these groups, while Miras 
(GCVS) amount to 45\,\% for Group 2 and 65\,\% for Group 3. 
All of this indicates that, in each population, the sequence of SRb 
Semiregulars corresponds to an evolutionary sequence towards the Mira 
instability strip.

\subsection{Projection onto the \{$P$, $V-K$\} plane}

In the same way as in the preceding subsection, a linear fit to the  
projected model distributions (see Fig. 6) yields the following 
period--colour relations: 
 
\begin{itemize} 
\item {Group 1}:  
$$(V-K)_0 = 10.80_{[\pm 0.58]} (\log P-2.48) + 8.51_{[\pm 0.05]} $$ 
 
\item {Group 2}:  
$$(V-K)_0 = 1.80_{[\pm 0.32]} (\log P-2.00) + 5.75_{[\pm 0.10]}$$ 
 
\item {Group 3}:  
$$(V-K)_0 = 2.54_{[\pm 0.28]} (\log P-1.75) + 6.20_{[\pm 0.10]} $$ 
 
\item {Group 4}:  
$$(V-K)_0 = 6.50_{[\pm 2.47]} (\log P-2.24) + 5.52_{[\pm 0.15]} $$ 
\end{itemize} 
 
The difference of slope between Groups 2 and 3 is (qualitatively) consistent 
with the differences of mass and metallicity expected from the kinematics. 
Indeed, the lar\-ger mass of Group 3 stars yields significantly higher 
temperatures and thus a larger \{$T_{\rm eff},(V-K)$\} slope, while the 
moderate metallicity difference has only a small influence (Bessell et al. 
\cite{bessell89, bessell98}). This is due to the behaviour of the TiO lines in 
this temperature range. 

On the other hand, the much larger slope of Group 1 may be explained by 
a difference of pulsation mode, consistently with the larger mean period. 
Indeed, the period of a lower--order mode must be more sensitive to the 
temperature (see, e.g., Barth\`es \cite{barthes98}).

\begin{figure} 
\vspace{3in} 
\includegraphics{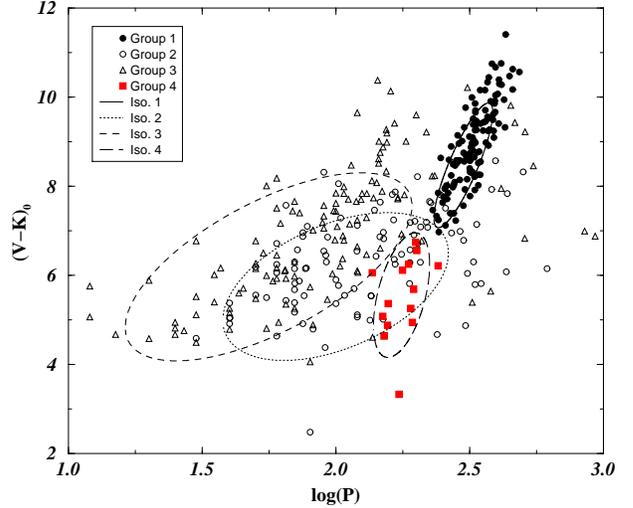} 
\caption[]{PC calibrated distributions: individual data and 
projected model distributions.} 
\label{fig6} 
\end{figure}

\subsection{Projection onto the \{$M_K$, $V-K$\} plane} 

The calibration results in the Luminosity--Colour plane are shown in Fig. 7. 
As explained in the Appendix A, the offset and the difference of width 
between the data distributions and the projected $2\sigma$ contours are 
effects of the projection and of the sampling bias. 
A linear fit to the contours yields the following luminosity--colour 
relations: 
 
\begin{itemize} 
\item {Group 1}:  
$$M_K = -0.42_{[\pm 0.05]} ((V-K)_0-8.47) -7.24_{[\pm 0.05]} $$ 
 
\item {Group 2}:  
$$M_K = -0.45_{[\pm 0.03]} ((V-K)_0-5.75) -6.41_{[\pm 0.03]} $$ 
 
\item {Group 3}:  
$$M_K = -0.34_{[\pm 0.04]} ((V-K)_0-6.19) -6.51_{[\pm 0.06]} $$ 
 
\item {Group 4}:  
$$M_K = -0.10_{[\pm 0.17]} ((V-K)_0-5.56) -7.06_{[\pm 0.14]} $$ 
\end{itemize} 

As in the preceding subsection, the difference of slope between Groups 2 and 
3 is easily explained by the tempe\-rature--dependence of the 
\{$T_{\rm eff}, (V-K)$\} slope, which is little sensitive to moderate 
metallicity variations.

\begin{figure} 
\vspace{3in} 
\includegraphics{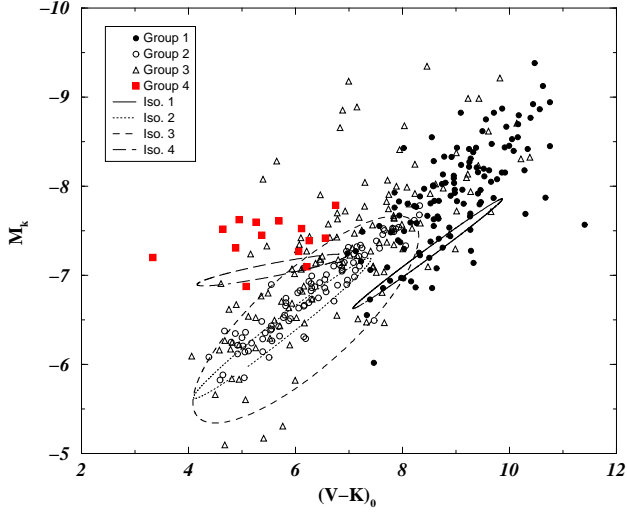} 
\caption[]{LC calibrated distributions: individual data and 
projected model distributions.} 
\label{fig7} 
\end{figure}

\subsection{\{P, $J-K$\} distribution}

Once the distances have been calibrated, it is possible to check the  
distribution of the sample stars with respect to the de-reddened $J-K$  
index. The results are shown in Fig. 8, together with the raw data.  
 
The $J-K$ Period--Colour distribution appears similar to the $V-K$ one.  
The scattering, also existing in the raw data, makes the PC relation more  
difficult to see. It is due to the smaller number of stars in the $J$ data 
set, and probably also to the peculiar sensitivity of this colour index to 
the surface gravity and extension of the envelope (Bessell et al. 
\cite{bessell89, bessell98}). 

A linear least--squares fit to the de-reddened data (excluding a few 
obviously misclassified stars, namely two having $(J-K)_0 > 2$ and two having  
$(J-K)_0 < 0.6$) yields: 
 
\begin{itemize} 
\item {Group 1} (86 stars):  
$$(J-K)_0 = 1.01_{[\pm 0.22]} \log P - 1.27_{[\pm 0.56]}$$   
$$\sigma_{J-K}=0.02$$  
 
\item{Group 2} (63 stars): 
$$(J-K)_0 = 0.16_{[\pm 0.07]} \log P + 0.83_{[\pm 0.14]}$$   
$$\sigma_{J-K}=0.02$$  
 
\item{Group 3} (91 stars): 
 $$(J-K)_0 = 0.17_{[\pm 0.04]} \log P + 0.85_{[\pm 0.08]}$$  
$$\sigma_{J-K}=0.02$$  
 
\item{Group 4} (12 stars): 
 $$(J-K)_0 = 1.02_{[\pm 0.67]} \log P - 1.25_{[\pm 1.51]}$$ 
$$\sigma_{J-K}=0.01$$  

\end{itemize} 
 
These fit relations are probably slightly biased, and thus should be 
shifted by a certain amount so as to represent the whole populations. 

Contrary to what was found with $V-K$, the relations of Groups 2 and 3 
cannot be reliably distinguished here. This may be due to the crossing--over 
of the iso--metallicity curves in the \{$T_{\rm eff}, J-K$\} diagram: the 
effects of the differences of temperature and metallicity between the two 
groups tend to compensate each--other (Bessell et al. \cite{bessell89, 
bessell98}).\\ 

\begin{figure}[htbp] 
\vspace{15cm} 
\includegraphics{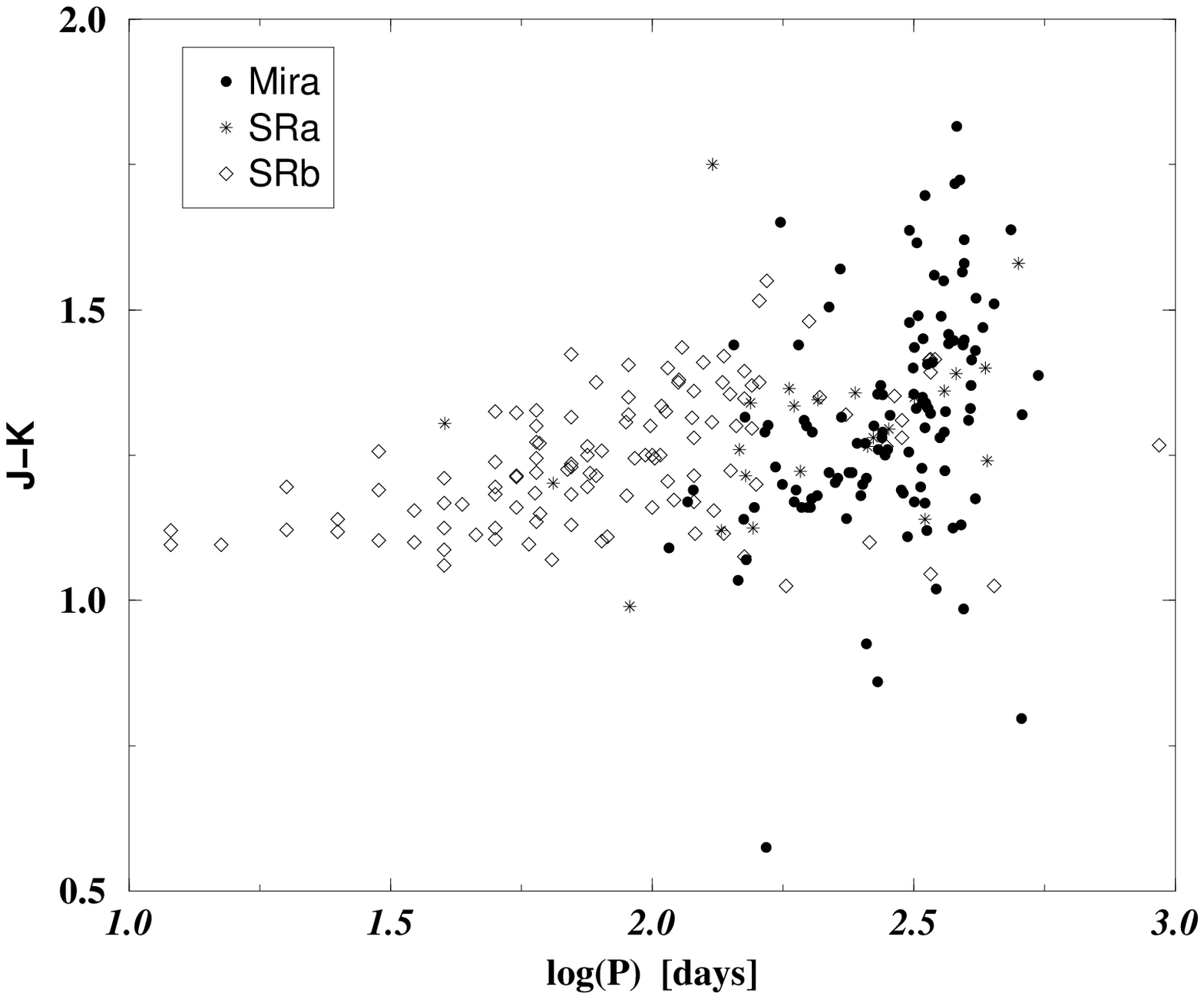} 
\includegraphics{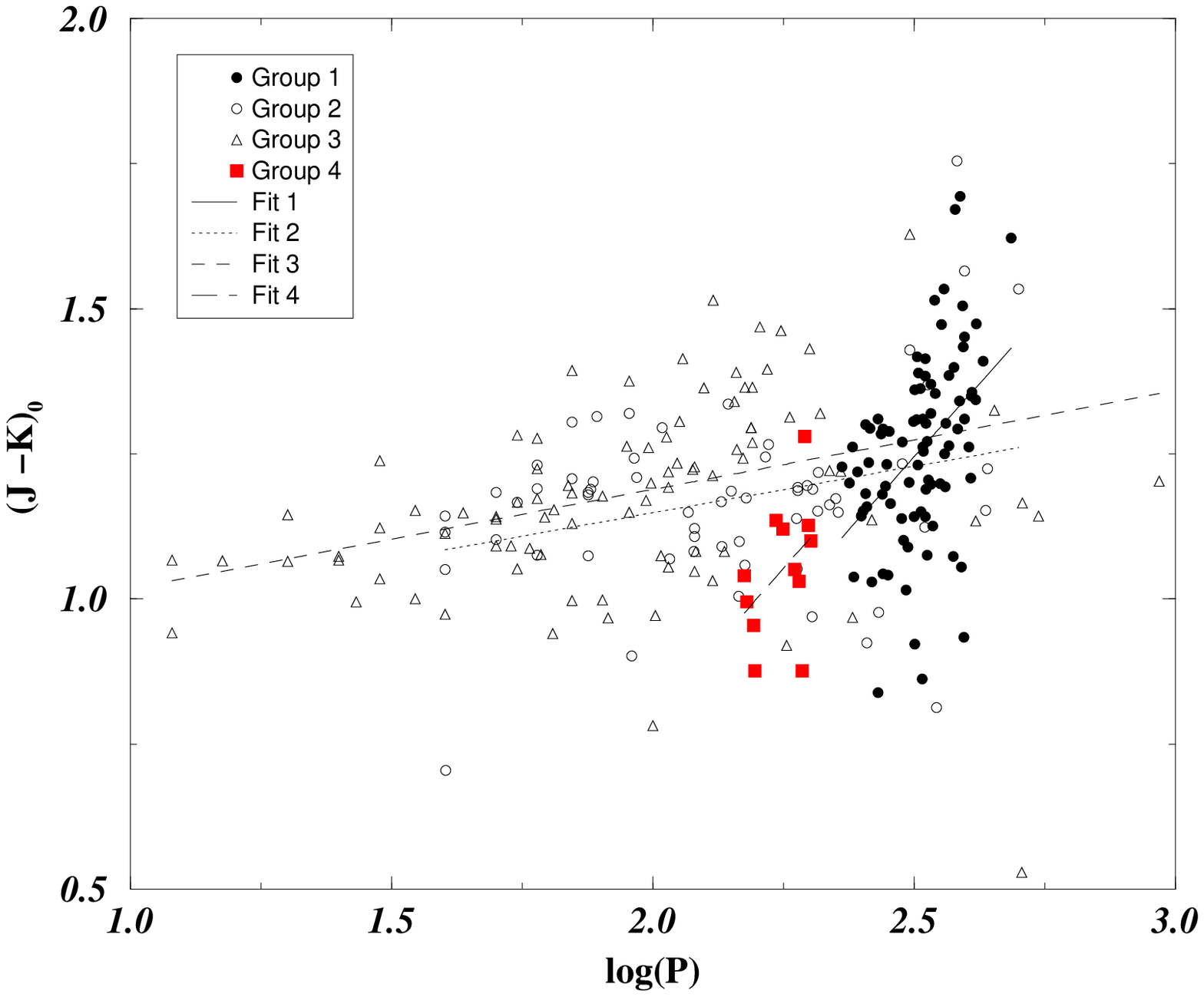} 
\caption[]{$J-K$ versus Period distribution of sample stars: raw ({\it top}) 
and dereddened ({\it bottom}) indices. Two stars with $J-K > 2$ are not 
shown.} 
\label{fig8} 
\end{figure}

Based on 29 Oxygen Miras, the relation found by Feast et al. (\cite{feast89}) 
for the LMC is:  
 
$$(J-K)_0 = 0.56_{[\pm 0.12]} \log P - 0.12_{[\pm 0.29]}$$  
$$\sigma_{J-K}=0.08$$  
 
From a sample of 21 stars, Hughes (\cite{hughes93}) derived:  
 
$$(J-K)_0 = 0.37_{[\pm 0.05]} (\log P - 2.4) + 1.215_{[\pm 0.014]}$$  
$$\sigma_{J-K}=0.06$$ 
 
As for the PL relation, we find a significant discrepancy between the Miras  
in the LMC and the ones in the solar neighbourhood. Since, in $J-K$ as well 
as in $V-K$, Group 4 is approximately aligned with Group 1, and thus  
metal--deficient Miras of the LMC should not significantly influence its PC 
relation, it seems that, as suggested in Sect. 5.2, we are actually 
encountering a problem of misclassification of the LPVs in the outer 
galaxies. This, of course, does not preclude the existence of a 
metal--deficient population which would further influence the PL relation.

\subsection{\{$M_K$, $J-K$\} distribution} 

The LC relations yielded by a linear least--squares fit to the calibrated 
and de-reddened data are: 

\begin{itemize} 
\item {Group 1}:  
$$M_K = -1.22_{[\pm 0.39]} (J-K)_0 -6.32_{[\pm 0.49]} $$ 
$$\sigma_M=0.33$$ 
 
\item {Group 2}:  
$$M_K = -1.37_{[\pm 0.37]} (J-K)_0 -5.24_{[\pm 0.44]} $$ 
$$\sigma_M=0.22$$ 
 
\item {Group 3}:  
$$M_K = -2.39_{[\pm 0.59]} (J-K)_0 -4.41_{[\pm 0.70]} $$ 
$$\sigma_M=0.69$$ 
 
\item {Group 4}:  
$$M_K = -0.18_{[\pm 0.65]} (J-K)_0 -7.25_{[\pm 0.68]} $$ 
$$\sigma_M=0.06$$ 
\end{itemize} 

We remind that these fit relations are subject to sampling bias, and thus 
should be significantly shifted downwards, so as to represent the whole 
population (see Appendix A). 
 
In view of the error bars, the slope of Group 3 may be the same as the  
one of Groups 1 and 2, which is what is expected from AGB evolutionary  
models. This supports our interpretation of the slope differences found  
in \mbox{$V-K$} (Sect. 5.4). The slope of Group 4 is, once again, not 
reliable.

\begin{figure} 
\vspace{3in} 
\includegraphics{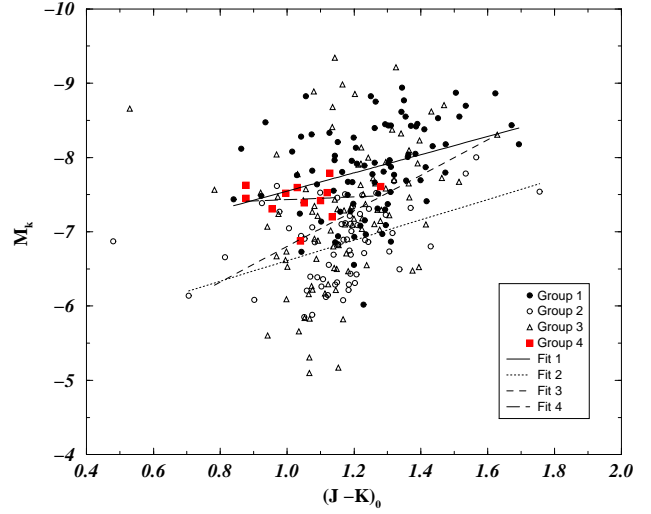} 
\caption{Magnitude versus $(J-K)_0$ distribution of the sample stars, 
deduced from the luminosity calibration.} 
\label{fig9} 
\end{figure}

\section{Conclusions} 
 
Thanks to an up--to--date maximum--likelihood method, using parallaxes, 
proper motions, radial velocities and other independent data (periods 
and colours), we have calibrated the luminosity of about 350 Oxygen--rich  
Long--Period Variable stars (Miras, SRa and SRb) observed by {\sc Hipparcos}.  
Meanwhile, the stars got classified in several groups differing by their  
distributions of kinematical parameters, magnitude, period and colour.  
Four groups were found: Group 1, mainly composed of Miras and having the  
kinematics of old disk stars; Group 2, mainly composed of SRb, with the  
same kinematics as Group 1; Group 3, also mainly composed of SRb, but with  
a much younger kinematics; Group 4, corresponding to the extended disk and  
halo, and containing no SRb. For each of them, we obtained de--biased PLC,  
PL, PC and LC relations.\\  
 
We thus confirm the existence of two SRb populations, already suggested on  
other grounds by Kerschbaum \& Hron (\cite{kersch92, kersch94}).  
 
SRa stars do not seem to constitute a separate class. They can be considered  
as a small--amplitude subset of the Mira and SRb classes. 
 
The presence of a small but significant number of Miras in Group 2, 
and of SRb's in Group 1, may be due to the probabilistic character of our 
classification method. However, it also tends to confirm that the usual (GCVS) 
classification criteria are not fully pertinent, as shown by Lebzelter et al. 
(\cite{lebz95}). 

As expected, since they all belong to the AGB, the stars seem to obey a  
global luminosity--colour relation, both in $(V-K)_0$ and $(J-K)_0$.  
More precisely, each group has its own relation, nearly parallel to the  
others, with a slight shift. 
 
Though belonging to the same Galactic population, Group 2 stars (SRb's) are 
fainter, bluer, and have shorter period and shallower period--colour relation 
than Group 1 (Miras). They probably pulsate on a higher--order mode. 
The higher luminosity and shorter period of Group 3 with respect to Group 2 
is probably due to higher mass and metallicity. The slope of Groups 2 and 3 
in a period--luminosity diagram, as well as a close look at the distribution 
of the three variability types within these groups, indicate that the two 
sequences of Semiregulars correspond to evolutionary sequences towards the 
Mira instability strip (i.e. SRb's are, generally, a little younger than 
Miras of the same population). 

All these findings will be confronted in detail to theoretical models of  
pulsation and evolution in Paper II.\\ 
 
Another important result of our study is that, contrary to a usual assumption 
(e.g. in van Leeuwen et al. [\cite{vanleeuw97}]), but consistently with the 
work of Menzies \& Whitelock (\cite{menzies85}) on a few Globular Cluster 
stars, the PL and PC relations of Oxygen--rich Miras found in the LMC 
{\it may not} be trivially transposed to other galaxies by simply shifting 
the zero--point: their slopes are inconsistent with the ones found for 
O--rich Miras in the solar neighbourhood. 

The first explanation is a misclassification of many LMC stars, since the 
observers simply discriminated the SRa stars on grounds of their small 
amplitude, and thus the remaining so--called ``Miras'' included a significant 
proportion of (younger) SRb stars. Concerning the PL relations, 
additional discrepancy may be generated by a metal--deficient, probably older 
LMC Mira population, the existence of which was already suspected by Wood et 
al. (\cite{wood85}) and Hughes et al. (\cite{hughes91}). All this, together 
with the fact that the stars distribution within the LMC ``Mira'' strip (at 
least below 500 days) seems rather uniform, suggests that these LPVs derive 
from a quite smooth star formation history, rather than well--separated 
bursts.\\ 

Concluding, the fact that the global PL relation of LMC ``Miras'' 
approximately matches the global one of Galactic Miras (in the solar 
neighbourhood) does not guarantee that it holds for every galaxy: everything 
depends on the relative number of misclassified SRb's and on the respective 
proportion of the different populations of stars, i.e. on the star formation 
history. 

Nevertheless, consistently with the LMC and globular clusters data (see e.g. 
Hughes \& Wood \cite{hughes90} and Menzies \& Whitelock \cite{menzies85}), 
our calibrations show that an LPV \mbox{M--giant} (Mira, SRa or SRb) 
pulsating with a period of 300--330 days is expected to have a mean absolute 
$K$ magnitude of $-7.5 \pm 0.5$, {\it whatever the stellar population}. 
This may be used as a distance estimator.\\

\medskip
 
\acknowledgements{This work was supported by the European 
Space Agency (ADM-H/vp/922) and by the hispano--french Projet 
International de Coop\'eration Scientifique (PICS) N$^o 348$. 
R.A. benefits from an EU TMR ``Marie Curie'' Fellowship. We also gratefully 
thank J.A. Mattei and the AAVSO staff for their help in evaluating the GCVS 
data.} 

\appendix 

\section{Populations, sampling and biases}

A striking feature of Figs. 5 through 7 is that many stars are located 
outside, most often above the projected $2\sigma$ contour of the fitted 
distribution of the corresponding group, whereas one would expect most of 
them to be located inside or symmetrically around it. 

First, one must remember that, in the PL and LC diagrams, the projected 
$2\sigma$ contour (i.e. the projection of the $2\sigma$ contour of the mean 
PLC plane) {\it is not} the $2\sigma$ contour of the projected 
distribution. This explains why, in the Luminosity--Colour diagram 
(Fig. 7), the width of each sequence of sample stars is much larger than the 
corresponding ``ellipse''. 

On the other hand, the offset of the sample distributions with respect to 
the model ones is due to the fact that the sample selection is based on the 
apparent magnitude (see Sec. 3.1). This is analogous to the Malmquist bias. 
To better understand the phenomenon in our case, a closer look at this well 
known bias may be helpful:\\ 

Malmquist (\cite{malm36}) studied the bias in the mean absolute magnitude 
that is derived from a sample of stars with the following characteristics: 
\begin{itemize}
\item The base population from which the sample is extracted has (a) a 
homogeneous spatial distribution and (b) a Gaussian distribution of absolute 
magnitudes $G(M_{0},\sigma _{M})$ 
\item The sample is selected within the base population by means of a 
limit--magnitude criterium: $ m\leq m_{lim} $ 
\end{itemize}
Under these conditions, Malmquist (\cite{malm36}) proved that the mean 
absolute magnitude of the sample $<M>$ differs from the mean absolute 
magnitude of the base population $M_{0}$ according to: 
$ <M>=M_{0}-1.38\; \sigma _{M}^{2}$\,. 

\noindent In other words, the stars in the sample are, on average, brighter
that the base population. The reason for this is that, because of the
apparent magnitude limit, brighter stars are over--represen\-ted in the 
sample: as they can be seen at longer distances, more of them are included.\\ 

In our case, the effects are more complicated (inhomogeneous spatial 
distribution, PLC relations and complex selection function) and also 
stronger. The PLC relations have, in some cases, large slopes and thus the 
groups may contain stars of very different absolute magnitudes. Like in the 
case of the Malmquist bias, brighter stars are favoured and thus 
over--represented in our sample. This favours, in turn, stars with long 
periods and large colour indices but also, at a given period and colour, 
stars located on the ``bright'' side of the main PLC plane. 

\begin{figure} 
\vspace{3in} 
\includegraphics{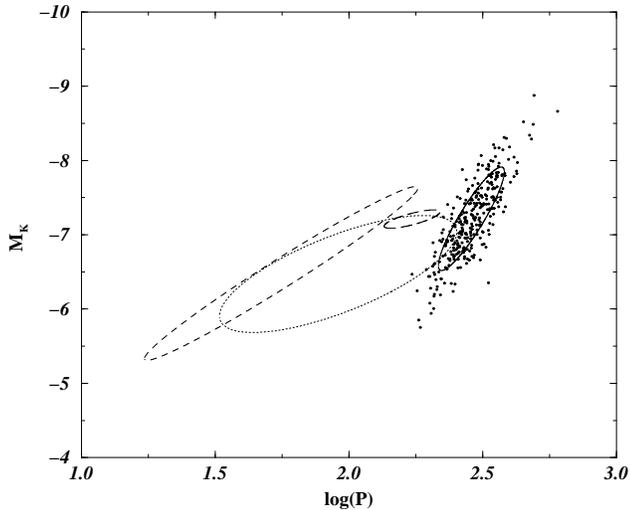} 
\caption{Simulated sample of Group 1 stars with no magnitude limit.} 
\label{figA1} 
\end{figure} 

\begin{figure} 
\vspace{3in} 
\includegraphics{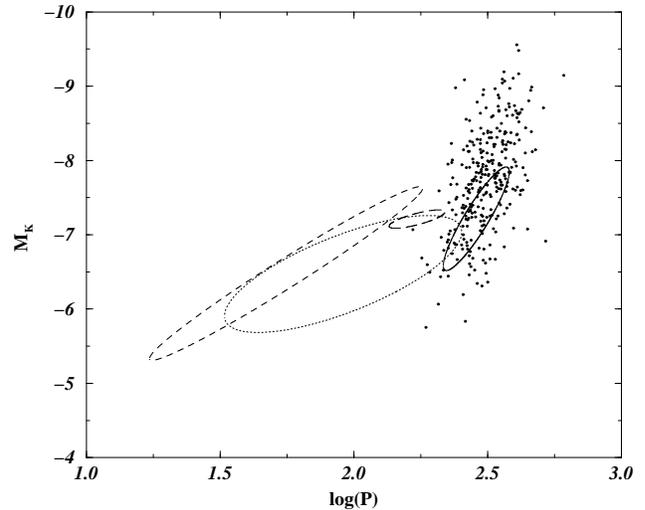} 
\caption{Same simulation when applying the magnitude--based selection 
function.}
\label{figA2} 
\end{figure} 

This effect can be illustrated by Monte-Carlo simulations. Let us first 
simulate a sample of Group 1 stars with no magnitude limit. As can be seen 
in Fig.~A1, most of these stars are located, in the  $\{\log P,M_{K}\}$ 
plane, inside the projected $2\sigma$ contour of the distribution used for 
the simulation. However, when the selection function (the one whose 
parameters were calculated in Sect. 4) is applied, the $\{\log P,M_{K}\}$ 
distribution of the sample drastically changes, as shown in Fig.~A2: in 
this case, a majority of the stars are brighter than the projected $2\sigma$ 
contour. 

This example clearly shows that the suprising peculiarities of Figs. 5 to 7 
are nothing but natural. Moreover it shows that, in the most general case, a 
``naive'' fit to the absolute magnitude distribution of a sample is not at all 
representative of the base population. Fortunately, this bias is probably 
negligible in the case of Magellanic Clouds studies, since all their LPVs may 
be considered as located at the same distance with a reasonable 
approximation, and their amplitudes are small in the near--infrared bands 
where they are observed.

\end{document}